\documentclass[a4paper,fleqn,usenatbib]{mnras}

\usepackage[T1]{fontenc}
\usepackage{ae,aecompl}

\usepackage{graphicx}	
\usepackage{amsmath}	
\usepackage{amssymb}	
\usepackage{lscape}
\usepackage{lmodern}
\usepackage{caption}

\title[Hermus Stream Orbit Fit]{An Orbit Fit to Likely Hermus Stream Stars}

\author[C. Martin et al.]{
Charles Martin,$^{1}$\thanks{E-mail: martic13@rpi.edu}
Paul M. Amy,$^{1}$
Heidi Jo Newberg,$^{1}$
Siddhartha Shelton,$^{1}$ \newauthor 
Jeffrey L. Carlin,$^{1,2}$
Timothy~C. Beers$,^{3}$
Pavel Denissenkov,$^{4}$
and Benjamin A. Willett$^{1}$
\\
$^{1}$Department of Physics, Applied Physics and Astronomy, Rensselaer Polytechnic Institute, Troy, NY 12180, USA\\
$^{2}$LSST and Steward Observatory, 950 North Cherry Avenue, Tucson, AZ 85721, USA\\
$^{3}$Department  of  Physics  and  JINA  Center  for  the  Evolution  of  the  Elements,  University  of  Notre  Dame, 
225 Nieuwland Science Hall, Notre Dame, IN 46556, USA\\
$^{4}$Department of Physics \& Astronomy, University of Victoria, Victoria, BC, V8W3P6, Canada
}

\date{Accepted 2018 March 2. Received 2018 March 1; in original form 2017 November 11}

\pubyear{2017}

\begin{document}
\label{firstpage}
\pagerange{\pageref{firstpage}--\pageref{lastpage}}
\maketitle

\begin{abstract}
We selected blue horizontal branch (BHB) stars within the expected distance range and sky position of the Hermus Stream 
from Data Release 10 of the Sloan Digital Sky Survey.  
We identify a moving group of $19$ BHB 
stars that are concentrated within two degrees of the Hermus Stream, between $10$ and $14$ kpc from the Sun. The concentration in velocity is inconsistent with a Gaussian distribution with 98\% confidence (2.33 sigma).
The stars in the moving group have line-of-sight velocities of v$_{\rm gsr} \sim 50$ km s$^{-1}$, a velocity dispersion of $\sigma_v\lesssim11$ km s$^{-1}$, a line-of-sight depth of $\sim 1$ kpc, and a metallicity of [Fe/H] $=-2.1 \pm 0.4$. The best-fit orbit has a perigalacticon of $\sim4$ kpc, apogalacticon of $\sim 17$ kpc, orbital period of $\sim247$ Myr, eccentricity $e = 0.62$, and inclination $i \sim75^\circ$ from $b=90^\circ$. The BHB stars in the stream are estimated to be $12$ Gyr old. An \textit{N}-body simulation of a mass-follows-light ultrafaint dwarf galaxy with mass $10^6M_{\odot}$ and radius $40$ pc is consistent with the observed properties. The properties of the identified moving group of 19 BHB stars are close enough to those of the Hermus Stream (which is traced predominantly in turnoff stars) that we find it likely that they are associated. If that is the case, then our orbit fit would imply that there is no relationship between the Hermus and Phoenix streams, as previously proposed.
\end{abstract}


\section{Introduction}

The halo of the Milky Way (MW) is full of tidal debris streams and clouds left from smaller galaxies that have been accreted over its lifetime \citep{2013NewAR..57..100B, 2016ASSL..420...87G}. These streams and clouds can be used as constraints for \textit{N}-body simulations when calculating the mass or potential of the MW \citep{2005ApJ...619..807L, 2010ApJ...711...32N, 2010ApJ...712..260K, 2014MNRAS.445.3788G}. They can also be used to explore the accretion history of the stellar spheroid in the MW \citep{1999MNRAS.307..495H} or galaxies in general \citep{2016arXiv161205471C}.

Although tidal streams that could be identified by density contrast were once thought to be non-existent or rare, a large number of streams have now been identified. Known tidal streams include the Sagittarius stream \citep{2001ApJ...547L.133I, 2003ApJ...599.1082M, 2006ApJ...642L.137B}, GD-1 Stream \citep{2006ApJ...643L..17G}, Orphan Stream \citep{2007ApJ...658..337B, 2010ApJ...711...32N, 2013ApJ...776...26S}, Virgo Stellar Stream \citep{2001ApJ...554L..33V, 2007ApJ...668..221N, 2006ApJ...636L..97D, 2014A&A...566A.118D}, Cetus Polar Stream \citep{2009ApJ...700L..61N, 2013ApJ...776..133Y}, Pisces Stellar Stream \citep{2012ApJ...760L...6B, 2013ApJ...765L..39M}, Alpheus \citep{2013ApJ...769L..23G}, PAndAS MW stream \citep{2014ApJ...787...19M}, ATLAS stream \citep{2014MNRAS.442L..85K}, Ophiuchus stellar stream \citep{2014MNRAS.443L..84B}, and Phoenix Stream \citep{2016ApJ...820...58B}. These streams are coherent spatial structures that are formed from dwarf galaxies or globular clusters that were stripped apart by tidal forces in the MW. Stellar clouds, like the Virgo Overdensity \citep{2001ApJ...554L..33V, 2002ApJ...569..245N, 2012ApJ...753..145C}, Pisces Overdensity \citep{2007AJ....134.2236S, 2009ApJ...705L.158K, 2010ApJ...717..133S, 2011RMxAC..40..261V}, Hercules-Aquila Cloud \citep{2007ApJ...657L..89B, 2014MNRAS.440..161S}, and Eridanus-Phoenix Cloud \citep{2016ApJ...817..135L}, are stellar overdensities spread over a large spatial area. See \citet{2016ASSL..420.....N} for a review of tidal streams and substructures. Unfortunately, no review is up-to-date, since new stellar streams in the halo continue to be discovered \citep{2016MNRAS.463.1759B, 2017ApJ...834...98G, 2018arXiv180103097S}.  

Many studies of the shape of the MW's potential have focused on the Sagittarius dwarf tidal stream \citep{2001ApJ...551..294I, 2005ApJ...619..800J, 2010ApJ...714..229L, 2012ApJ...744...25C, 2013ApJ...773L...4V, 2016MNRAS.458L..64G, 2017arXiv170302137D}, because it is the largest stream with the most information. Others have concentrated on narrow globular cluster streams like Pal 5 and GD-1 \citep{2009ApJ...697..207W, 2010ApJ...712..260K, 2010PhDT.......194W, 2016ApJ...833...31B}, which are simpler because these systems have less dependence on the progenitor properties. The narrower globular cluster streams are also used to study the gaps expected to form in tidal streams when subhalos pass through them \citep{2009ApJ...705L.223C, 2013ApJ...768..171C, 2016PhRvL.116l1301B, 2016MNRAS.463..102E}. The results of these studies have been interesting, but unfortunately they do not yet yield a consistent picture of the MW halo.

There has been considerable recent interest in methods for constraining the Galactic potential using tidal streams, particularly since the Gaia satellite is expected to deliver spatial and kinematic information for many tidal streams \citep{2013ApJ...778L..12P, 2014ApJ...795...94B, 2015ApJ...801...98S, 2017ApJ...836..234S}. The first data release has already been used to explore the substructure of the stellar halo close to the Sun \citep{2017A&A...598A..58H}.

The problem of using tidal streams to determine the distribution of dark matter in the MW is complex. The spatial density distribution in the MW could be arbitrarily complex and time-dependent. In addition, the tidal streams we use as tracers may have their own complex structures and merger histories, or fall into the MW potential as groups. However, by using many well-studied streams with documented orbital parameters, it might be possible to disentangle all of the unknowns due to the tremendous amount of data provided by the tidal disruption of satellites. Early attempts relied on orbit-fitting, and were subsequently called into question because tidal streams only approximately follow the orbital path of the progenitor satellite \citep{2013MNRAS.433.1813S}. New methods are being developed that address these issues.

To ascertain the MW potential, it is not only important to discover new tidal debris streams, it is also important to refine our understanding of existing streams as newer data becomes available. In particular, it is vitally important to know that the data for each tidal stream is accurate and not a product of multiple tidal streams. The misidentification of tidal streams has been surprisingly common. For example, the BHBs that were associated with the original detection of the Sgr stream in the south Galactic cap \citep{2002ApJ...569..245N} turned out to be primarily members of the Cetus Polar Stream \citep{2009ApJ...700L..61N}. It has even been suggested that the Cetus Polar Stream might itself be composed of two or more streams \citep{2012ASPC..458..219G}.

In this paper, we will examine the Hermus stream, which along with the Hyllus stream was discovered by \citet{2014ApJ...790L..10G} (hereafter G14). This stream was discovered using the Sloan Digital Sky Survey \citep[SDSS;][]{2000AJ....120.1579Y} photometry using a matched-filter technique that was pioneered with the discovery of the Pal 5 tidal stream \citep{2001ApJ...548L.165O, 2002AJ....124..349R}, but was extended by Carl Grillmair to find many low surface brightness halo substructures: the NGC 5466 tidal tail \citep{2006ApJ...639L..17G}; GD-1 \citep{2006ApJ...643L..17G}; an independent detection of the Orphan stream \citep{2006ApJ...645L..37G}; the anti-center stream and EBS \citep{2006ApJ...651L..29G}; and Acheron, Cocytos, Lethe and Styx \citep{2009ApJ...693.1118G}. 

Although the matched-filter technique has been productive in identifying faint tidal streams in the Milky Way halo, the distances estimated in the discovery papers are fairly approximate. For example, \cite{2006ApJ...645L..37G} estimated a distance to the Orphan Stream of 21 kpc, with an orientation perpendicular to our line of sight. The orbit actually ranges from 19 to 47 kpc from the Sun \citep{2010ApJ...711...32N} over the detected extent of the stream. Similarly, \citet{2012ApJ...760L...6B} estimated the distance to the Pisces Stellar Stream to be $26 \pm 4$ kpc using isochrones with a metallicity of [Fe/H]$=-1.0$. In \citet{2013ApJ...765L..39M}, using the SDSS spectroscopic metallicity of [Fe/H]$=-2.2$ for giant branch members, the actual distance to the stream is closer to $35$ kpc.

In this paper, we identify a moving group of blue horizontal branch (BHB) stars that are spatially associated with the Hermus Stream, and fit an orbit to them. Likely BHB stars selected from the SDSS and SEGUE survey \citep{2009AJ....137.4377Y}, within $\pm 2^\circ$ of the Hermus Stream orbit, show an overdensity at $12$ kpc and v$_{\rm gsr}\sim50$ km s$^{-1}$.  The statistical significance of this moving group is explored with several tests, and we conclude that it is a coherent velocity substructure with 98\% confidence.

Eleven of these stars trace a narrow orbit with a FWHM of $1.^\circ6$, metallicity of [Fe/H] $=-2.1$, and a distance of $12$ kpc from the Sun. Our orbit fit is defined by $(X,Y,Z) = (-2.3, 5.7, 8.7)$ kpc and velocity of $\mathbf{v}_0= (-123, 155, 59)~\text{km}~\text{s}^{-1}$.

G14 found the Hermus Stream to extend from a heliocentric distance of $15\pm3$ kpc at $l\sim70^\circ$ to a distance of $20\pm3$ kpc at $l\sim15^\circ$, with a full width of $0.^\circ 7$, and a metallicity of [Fe/H]$\sim-2.3$. The low metallicity is consistent with our identification of a tidal stream that is rich in BHB stars with metallicity of [Fe/H] $= -2.0 \pm 0.4$. However, our distance is closer than that measured in G14, and apparently outside of the error bars at low Galactic longitudes.  Despite this, we argue that the properties of the moving group are close enough to those of the Hermus Stream that they are likely associated. If so, it appears that the sky position of the southern portion of the Hermus Stream (near $l\sim15^\circ$) was simply misidentified, as was previously suggested in \citet[][hereafter G16]{2016ApJ...820L..27G}.

G16 suggested that the Hermus Stream was significantly farther away ($20$ kpc over the observed portion) than the original estimate for G14, and therefore an orbit could be constructed that associated it with the Phoenix Stream. This was attractive because both streams are cold and oriented in a similar plane; there was no other reason given for the more distant estimate for the Hermus Stream. Because the stars we identify in this stream are considerably {\it closer} than assumed by G16, our moving group is not consistent with a link between the Hermus Stream and the Phoenix Stream. This result underscores the need to verify which stars are associated with which streams before using them to draw conclusions about the accretion of the stellar halo and the distribution of dark matter in the MW.

This paper is structured as follows. Section 2 describes the selection of BHB stars in an On and Off field along the Hermus Stream. Section 3 addresses the distribution of these BHB stars and selection of Hermus Stream candidates. Section 4 proposes a new orbit for the Hermus Stream based on these BHBs. Section 5 shows that candidate Hermus Stream BHBs have a narrow age distribution. Section 6 demonstrates that an \textit{N}-body simulation of a ultra faint dwarf galaxy matches the spatial distribution of Hermus Stream BHBs. Section 7 rules out any association with our Hermus Stream orbit with other known MW substructure. Our conclusions are presented in Section 8.

\section{Data Selection}

\begin{figure}
\includegraphics[width=6in]{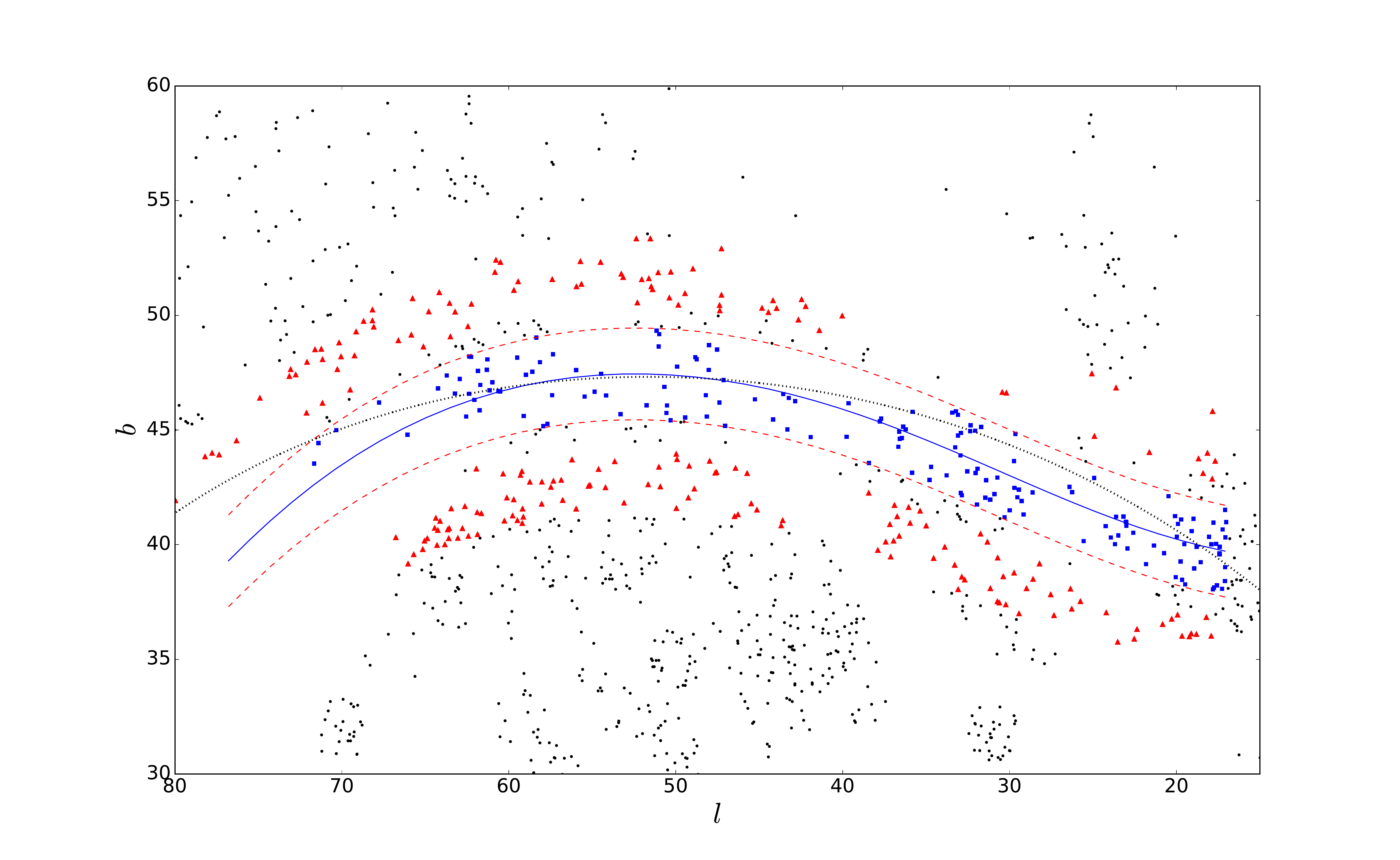}
\caption{Distribution of likely BHB stars with SDSS spectra (black points): $-0.25<(g-r)_0< 0$, $0.8<(u-g)_0<1.5$, and $0.0< \log g_{WBG} < 3.5$. The solid blue line represents the third-order polynomial fit to the Hermus Stream, as calculated by G14, while the black dotted line represents the best-fit orbit found in this work. We select an On field of stars to be within $2^\circ$ of the stream (blue squares) and an Off field (red triangles) to be between $3^\circ$ and $6^\circ$ of the stream. 
}
\label{Hermus}
\end{figure}

Figure \ref{Hermus} shows the distribution of likely BHBs with available medium-resolution ($R\sim 1800$) spectra from SDSS Data Release 10 \citep[DR10;][]{2014ApJS..211...17A} in the region of the Hermus Stream, selected using $-0.25<(g-r)_0< 0$, $0.8<(u-g)_0<1.5$ \citep{2000ApJ...540..825Y}, and $0.0< \log g_{WBG} < 3.5$. The WBG \citep{1999AJ....117.2329W} surface gravities were used because they are a better measure for stars with $(g-r)_0 < 0.25$ \citep{2009ApJ...700L..61N}. 

To explore the velocity, distance, and metallicity of the Hermus Stream, we select BHBs within $2^\circ$ Galactic latitude (hereafter the On field) of the third-order polynomial fit to the Hermus Stream as calculated by G14 (solid blue line in Figure \ref{Hermus}).  The width of the On field was chosen to be significantly wider than the published stream width, since the exact path of the stream is not well established; the discovery paper established that the stellar density was higher on the polynomial than it was adjacent to the polynomial, but did not establish that the density was higher along the entire length of the polynomial.   The fifth-order polynomial fit to the Hermus Stream described in G14 was not used because it did not result in the additional detection of likely Hermus Stream stars, and because it was rejected by G16. 

For comparison, we select BHBs in an off-stream part of the sky, between $3^\circ$ and $6^\circ$ from the third-order polynomial fit to the Hermus Stream (hereafter the Off field). Because the G14 estimate of the width of the Hermus Stream was $\sim 0.^\circ7$ wide, we expect that the Off regions should represent a background devoid of Hermus Stream candidates. However, there is no expectation that the Off regions should be devoid of other tidal debris.  In fact, they should at least contain stars in the Hyllus Stream, which was identified in the same G14 data.

\section{Distance and Velocity of BHB Candidates near the Hermus Stream}

In this section we study distance and velocity distributions of stars in the On and Off fields, within a distance range of $10$ to $25$ kpc. This is the distance range for Hermus Stream stars given in G14. We find one significant group of stars with similar distances and velocities in the On field, and not much of interest in the Off field.

\begin{figure}
\includegraphics[width=6.5in]{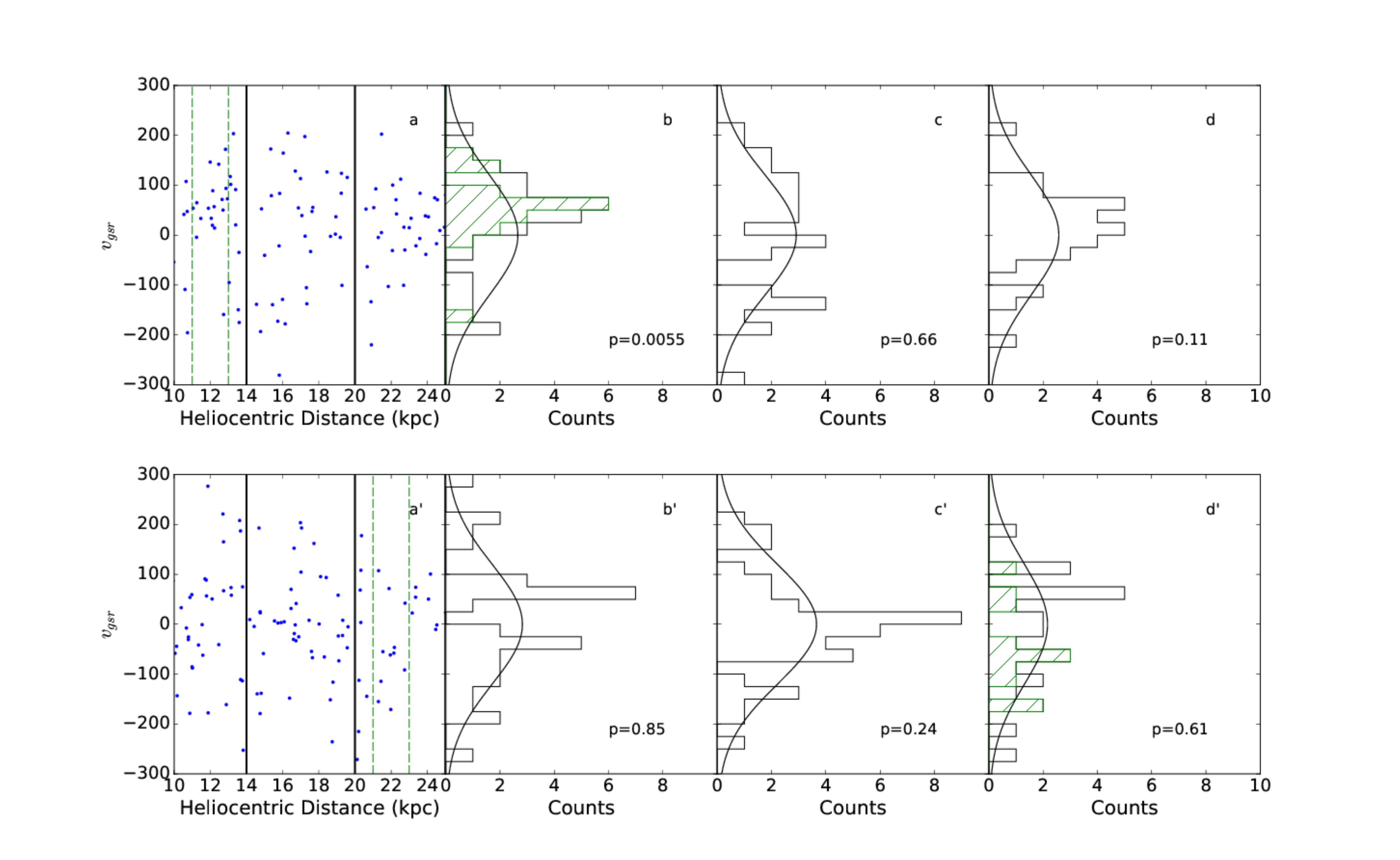}
\caption{{\it Top row:} On field BHBs within $2^\circ$ of Hermus Stream. The left panel shows v$_{\rm gsr}$ vs. distance, spanning the Hermus Stream range suggested in G14. We have chosen to split the selection at $d = 14$ kpc and  $d = 20$ kpc. The green dashed lines show the distance range for which the observed velocities are least consistent with being drawn from a Gaussian halo distribution. The remaining panels, from left to right, represent the velocity histograms for $10 < D < 14$ kpc, $14 < D < 20$ kpc, and $20 < D < 25$ kpc, respectively. Gaussians centered at zero velocity with a standard deviation of $120$ km s$^{-1}$ that have been normalized to the number of stars in each panel are shown to represent the expected distribution. The green hatched histogram shows the velocity distribution for stars between the green lines in the left panel. {\it Bottom row:} Off field BHBs $3^\circ$ - $6^\circ$ from Hermus Stream. The panels from left to right are the same as for the top row. Note the overdensity at (D, v$_{\rm gsr})=(12$ kpc$, 50$ kms$^{-1})$ in the On field, which we identify as the Hermus Stream.}
\label{vgsr_dist}
\end{figure}

The top and bottom rows of Figure \ref{vgsr_dist} show the Galactic standard-of-rest, line-of-sight velocities, v$_{\rm gsr}$, for the On and Off fields, respectively.  
The line-of-sight velocity was calculated using the Sun's motion from \citet{2009AJ....137.4377Y}:
\begin{equation}
v_{\rm gsr} = v_{\rm helio} + 10.1 \cos(b)\cos(l) + 224 \cos(b)\sin(l)+ 6.7 \sin(b).
\end{equation}
The distance to the BHB stars was calculated using the absolute magnitude from Equation 7 of \citet{2011MNRAS.416.2903D}:
\begin{equation}
\begin{split}
M_{g(BHB)} &=& 0.434 - 0.169(g-r)_0 + 2.319(g-r)^2_0 \\
  && {}+ 20.449(g-r)^3_0 + 94.517(g-r)^4_0,
\end{split}
\end{equation}
which is valid for stars with $-0.25 < (g-r)_0 < 0$.

In Figure 2 we identify an excess of positive velocity stars concentrated at distances between $10$ and $14$ kpc in the On field. 

Before measuring the significance of this moving group, it is important to review the context.  There is little room to adjust the parameters by which the BHB stars were selected.  The colors could be extended slightly to the blue and the red, but if we did that then our method for determining their distances would be invalid. The locations of the On and Off fields could be adjusted slightly, but not very much.  The center of the On field is set by the published position of the Hermus Stream, and the Off field is selected to be as near to the On field as possible so that it probes similar parts of the Milky Way.  The only freedom we have is the width of the sky that we select in the On field; the distance from the stream center could only sensibly be varied between about half a degree and a few degrees (six different possible values).

Our expectation is that if one selects a large portion of the stellar halo that the velocity distribution will be Gaussian with a sigma somewhere near 120 km s$^{-1}$ \citep{2005ApJ...622L..33B}.  We also know that this stellar component is likely to be built up from individual streams, and that if one chooses a small volume one is likely to find a moving group.  For example, \citet{1999MNRAS.307..495H} explored stars within 1 kpc of the Sun, which was not known to be in the neighborhood of a tidal stream, and found that 10\% of the halo stars had coherent motions.  This fraction decreased when a larger volume was surveyed \citep{2007AJ....134.1579K}. Between 20\% and 100\% of the stellar halo is thought to be built up from individual stellar streams \citep{2009ApJ...698..567S}.

The On and Off fields were then separated into samples with distances of $10<d<14$ kpc, $14<d<20$ kpc, and  $20<d<25$ kpc (see velocity histograms in Figure \ref{vgsr_dist}). These distance ranges were chosen to separate distance regions that appeared to have different character in the On field, but to not be so tightly constrained as to make a calculation of the internal statistics of the stars in each distance range meaningless. We will later show that a rigorous selection of all possible distance ranges gives similar probability results. 

A K-S test was performed to determine whether the velocity distribution in each distance sample (in both the On and Off fields) was consistent with a normal halo distribution with a mean line-of-sight velocity of zero and a standard deviation of $120$ km s$^{-1}$. The null hypothesis of these tests is that the population is drawn from a Gaussian distribution with a sigma of $120$ km s$^{-1}$; if the p-value is less than $0.05$, the dataset is inconsistent with the null hypothesis at the $95\%$ confidence level. We find that the closest stars in the On field are not consistent with being drawn from a Gaussian distribution ($p=0.0055$), while the more distant stars are ($p=0.66$ and $p=0.11$ for panels "c" and "d," respectively). Each Off field panel is consistent with a Gaussian distribution ($p=0.85$, $p=0.24$, and $p=0.61$ for panels "b$'$," "c$'$," and "d$'$" respectively). These results suggest that we might have found stars associated with the Hermus stream, at 10-14 kpc from the Sun.
We find no significant excess in the On field centered at $20$ kpc, the distance claimed for Hermus Stream by G16, within any velocity range.

Because we have separated the data into distance ranges by eye, we worried that the statistical validity of our results could be compromised by look-elsewhere effects \citep{2010EPJC...70..525G}.  This arises when one inspects a large number of parameters or data subsets, but only the unusual statistical cases are reported; with a large number of tests one would expect to find a few statistical flukes, and they would therefore not be significant.  In our case, we had the freedom to cut the data at any distance, and could have mentally tested many distance ranges before selecting distance ranges that would give the highest significance.

To test the effect of having the freedom to select an arbitrary distance interval, we ran a simulation in which we chose all possible distance subsets and then compared the p-value for each subset with the distribution of p-values in thousands of random samples of the same data. We performed K-S tests on all possible distance ranges of 2 kpc and larger, starting and ending on integer distances in kpc. The green dashed lines in the left panels of Figure \ref{vgsr_dist} represent the regions that are least like a Gaussian for both the On and Off fields. The stars within these regions are shown as green hatched histograms in the histogram panel that contains that distance range.

The regions least consistent with Gaussians in the On and Off fields have p-values of $p=0.0013$ and $p=0.064$, respectively. From running thousands of simulations of stars at the On field distances, but with velocities randomly sampled from a Gaussian distribution, we find that only $2\%$ of the time will the results yield a best p-value equal to or less than the $p=0.0013$ that was found for $11<d<13$ kpc. From this we conclude that the moving group is statistically significant at the 98\% confidence level.

\begin{figure}
\includegraphics[width=6in]{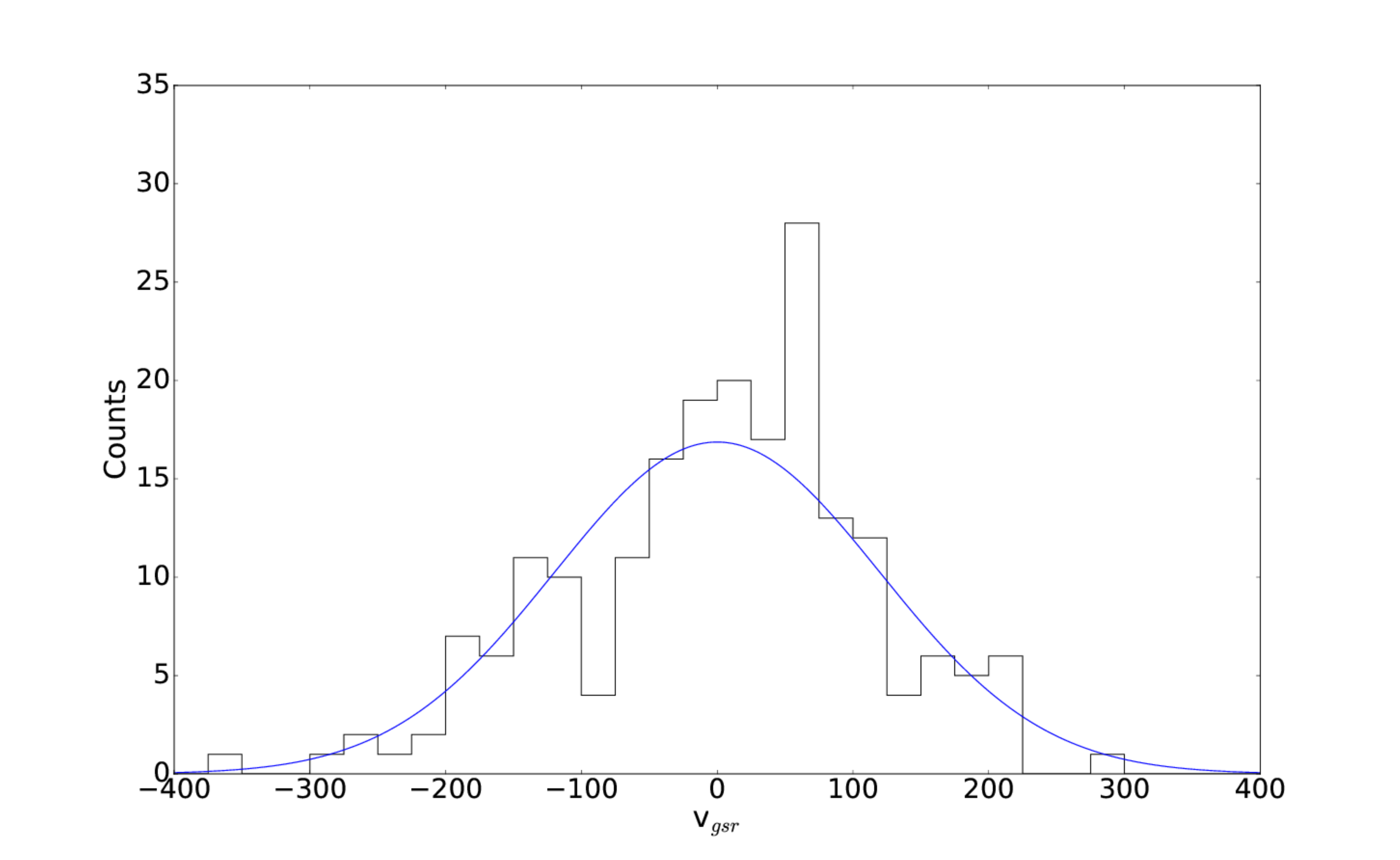}
\caption{V$_{\rm gsr}$ histogram of the total On and Off field stars from Figure \ref{vgsr_dist}. A Gaussian, centered at zero velocity with a standard deviation of $120$ km s$^{-1}$, normalized to the $203$ stars is included for reference. We can see that the distribution is roughly Gaussian excluding the excess at positive velocities associated with our structure.}
\label{New_vgsr_hist}
\end{figure}

In the previous statistical calculation, we used velocities sampled from a Gaussian because the spheroid population is well approximated by this distribution. Figure \ref{New_vgsr_hist} shows the velocities of all of the BHB stars with SDSS spectra in the On and Off fields between $10$ and $25$ kpc from the Sun. The distribution is reasonably Gaussian, with a significant deviation at $v_{gsr} \sim 50$ km s$^{-1}$ due to the identified moving group. 

In an abundance of caution, we also tested the probability of obtaining the observed moving group if the line-of-sight velocities were randomly assigned from the population (a permutation test) rather than from a Gaussian distribution. We randomly sampled velocities from the velocity histogram in Figure \ref{New_vgsr_hist} corresponding to each of the $105$ distances of stars in the top left panel of Figure \ref{vgsr_dist}, without replacement. This random sampling was repeated 15,000 times, and in each permutation we selected all distance ranges of 2 kpc or larger and performed a K-S test. In 15,000 random samples, we find that only $5\%$ of the time the best p-value is equal to or less than $\bf{p = 0.0013}$. We find a velocity substructure of p-value $0.0013$ or less twice as often using permutation as using a smooth Gaussian approximation to the distribution of halo star velocities, even though we did not remove the substructure itself from the random sample.  If only the Off field is used for the background distribution, then we find a structure as significant as our identified moving group only $1\%$ of the time.  Using the On field dilutes the significance because the stream is included in the velocity distribution, and using the Off field enhances the moving group because it likely contains other moving groups in other velocity ranges that are below our detection threshold.  We stand by the $98\%$ confidence level for detection of the moving group, which was obtained using a Gaussian prior for the distribution of line-of-sight halo velocities.

We select the 19 stars from the positive velocity structure, $10<$v$_{\rm gsr}<110$ km s$^{-1}$ and $10 < d < 14$ kpc as Hermus Stream BHB candidates. These stars are tabulated in Table \ref{tab:Table1}.

\section{Orbit Fit}
\subsection{Fitting Procedure}
We used the NEMO stellar dynamics toolbox \citep{1995ASPC...77..398T} to fit an orbit to the Hermus Stream BHB candidates. 
The best-fit was found by minimizing the Euclidean Mahalanobis distance \citep{Mahalanobis}, as developed by Benjamin Willett in Section 2.1.5 of his thesis \citep{2010PhDT.......194W}, and described in \citet{2009ApJ...697..207W}. This method optimizes the orbital parameters so as to minimize the $\chi^2$ deviation between the observations and the orbit fit, considering the line-of-sight velocity, position on the sky, and distance. We used SciPy's nonlinear conjugate gradient descent method to optimize the goodness-of-fit of NEMO-generated orbits to the observed BHB stars.

The goodness-of-fit function generates an orbit ($80$ Myr backward and forward from $l=45^\circ$ with a step size of $16,000$ years) and compares it with measured positions and velocities for the BHB stars. Each orbit is specified by values for $l, b, d, v_x, v_y,$ and $v_z$, in combination with a static gravitational potential composed of a Miyamoto-Nagai disk, a Hernquist bulge, and an NFW halo, with parameters described in Tables 1 and 2 of \citet{2015ApJ...811...36D}.

The first step is to search the orbit for the two $l$ values closest to that of each BHB star, one on each side. We then use a linear interpolation to obtain values for $b$, v$_{\rm gsr}$, and $d$:
\begin{eqnarray}
b_{\text{model,}i}=\frac{b_{\text{model,}k+1}-b_{\text{model,}k}}{l_{\text{model,}k+1}-l_{\text{model,}k}}(l_{\text{data,}i}-l_{\text{model,}k})+b_{\text{model,}k}; \nonumber \\ 
v_{\text{model,}i}=\frac{v_{\text{model,}k+1}-v_{\text{model,}k}}{l_{\text{model,}k+1}-l_{\text{model,}k}}(l_{\text{data,}i}-l_{\text{model,}k})+v_{\text{model,}k}; \nonumber \\
d_{\text{model,}i}=\frac{d_{\text{model,}k+1}-d_{\text{model,}k}}{l_{\text{model,}k+1}-l_{\text{model,}k}}(l_{\text{data,}i}-l_{\text{model,}k})+d_{\text{model,}k}.
\end{eqnarray}
Once these values are calculated for each data point, we calculate $\chi^2$ values for $b$, v$_{\rm gsr}$, and $d$:
\begin{eqnarray}
\chi_{\text{b}}^{2}=\sum_{i} \left(\frac{b_{\text{model,}i}-b_{\text{data,}i}}{\sigma_{\text{b}}} \right)^2 \nonumber \\
\chi_{\text{v}}^{2}=\sum_{i} \left(\frac{v_{\text{model,}i}-v_{\text{data,}i}}{\sigma_{\text{v}}} \right)^2 \nonumber \\
\chi_{\text{d}}^{2}=\sum_{i} \left(\frac{d_{\text{model,}i}-d_{\text{data,}i}}{\sigma_{\text{d}}} \right)^2,
\end{eqnarray}
then combine into a single fitness value:
\begin{eqnarray}
\chi_{\text{stream}}^2=\frac{1}{\eta}(\chi_{\text{b}}^2+\chi_{\text{v}}^2+\chi_{\text{d}}^2), 
\end{eqnarray}
where $\eta=N-n-1$.
Here \textit{N} is the number of inputs from the data points, and \textit{n} is the number of parameters to be fit. The $\chi^2$ value is the value to be optimized by the gradient descent, and the parameters to be fit are $b, d, v_x, v_y,$ and $v_z$. The parameter $l$ is fixed at $l=45^\circ$. The optimizer repeats the process of creating orbits; finding points; interpolating $b$, v$_{\rm gsr}$, and $d$; and calculating $\chi^2$; until it reaches a minimum value.

We initially used values of $\sigma_b=1^\circ$, $\sigma_v=15$ km s$^{-1}$, and $\sigma_r=1$ kpc for the uncertainties in the Galactic latitude, velocity, and distance, respectively. The Galactic latitude and velocity errors are the estimated width of the stream, while the distance error is the estimated measurement error in individual measurements.  After an orbit was fit to the BHBs, new standard deviations were calculated for latitude, velocity, and distance using the deviations from the orbit.

\subsection{Measurement Errors in SDSS}
To estimate the effect of measurement errors on our data, we searched through SDSS DR10 to find any duplicates of either stars in our dataset or duplicates of BHBs in the region of the Hermus Stream. We first made the same cuts in color index and surface gravity to remove non-BHBs, and then made cuts in position. In the first quadrant of the North Galactic Cap, at the distance and longitude of the Hermus Stream ($10$ kpc $\leq d \leq 14$ kpc; $17^\circ \leq l \leq
 72^\circ$), we found a total of nineteen stars with multiple measurements of radial velocity, eighteen of which were double measurements and one of which was a triple measurement. The average difference between measurements was $\Delta v_{r} = 10.37$ km s$^{-1}$ and the standard devation was 12.92 km s$^{-1}$. Five of these multiply-measured stars were also in our data set, and for those particular stars, the average difference between radial velocity measurements is 9.64 km s$^{-1}$ and the standard deviation is 6.87 km s$^{-1}$. This implies an inherent uncertainty of $\sim 10$ km s$^{-1}$ in velocity measurements.

Therefore, in order to account for uncertainties in the data, we randomized the distance $d$ and velocity v$_{gsr}$ within likely experimental error. For each data point, distance was allowed to randomly vary within 1 kpc of the calculated value and v$_{gsr}$ was allowed to vary within 10 km s$^{-1}$. We then performed 100 optimizations, each time with new randomized values. The best-fit orbit was then constructed by taking the average of the five optimized parameters, so that the orbit was described by $(l,\overline{b},\overline{d},\overline{v}_x,\overline{v}_y,\overline{v}_z)$. This method also allowed us to estimate the uncertainty in each fit parameter by calculating the dispersion of each.

\subsection{Effect of Interlopers}
In addition to the measurement errors, we can estimate how many of the nineteen stars in our dataset we expect to be interlopers from a background population. The Off field selection of panel a' of Figure 2 has a total of 11 stars within the velocity and distance selection used to select the candidate stream stars. This number is reduced to 8 stars when accounting for the area being ~33$\%$ larger than the On field. Examination of the negative velocity stars spanning the same range of distances yields a total of 3 stars in the On field and 9 stars in the Off field (which would be ~6 stars when accounting for the difference in area). The presence of substructure in the On field region would mean that the background population of BHB stars would be undersampled in that particular magnitude range; the presence of only one star in the negative velocity portion of the On field is explained by this effect. Using the numbers from the Off field, we can expect that we should see $\sim 7$ interloper stars in our selection of 19 stars, but the actual number might be lower due to background reduction in the region around halo substructures.

The interlopers also affect the velocity dispersion of the dataset. To estimate the effect of interlopers on the observed dispersion of the data, we generated 1000 artificial stars. The line-of-sight velocities of the created dataset were generated from a halo-like distribution: a Gaussian distribution with a mean of zero and a standard deviation of 120 km s$^{-1}$. The created stars were then cut using the Hermus selection criterion for velocity ($10$ km s$^{-1} \leq v_{\text{vgsr}} \leq 115$ km s$^{-1}$). This left us with roughly 300 stars, of which we randomly selected eight. The eight stars were then given Galactic longitude values consistent with that of the Hermus Stream ($17^\circ \leq l \leq 72^\circ$). We then combine these eight stars with eleven stars randomly selected from the actual data set (Table \ref{tab:Table1}) and calculate dispersions for both the eight generated stars and the eleven actual stars. We then combine these to see the effect of interlopers on the velocity dispersion:

\begin{eqnarray}
\sigma_{\text{comb}}^{2} = \frac{N\sigma_{\text{star}}^2+\sum\limits_{i=1}^{5}\left(v_{\text{gsr,}i}-\overline{v}_{\text{gsr}}\right)^2}{19}.
\end{eqnarray}

After ten thousand iterations, we find that the average variance of eleven randomly selected stars from the orbit is $\sigma_{\text{star}}^2 = 779.95$, while the average variance of the eight randomly generated points is $\sigma_{\text{gen}}^2 = 1404.26$. Combining these two using the above equation gives us $\sigma^2 = 1100.75$, which gives us an expected velocity dispersion of $\sim33$ km s$^{-1}$. This implies that the presence of interlopers in the original data set artificially inflates the velocity dispersion of the stars actually in the stream. 

To account for the effects of interlopers in the data on our orbit fits, we refitted the orbit from the initial point and with the initial dispersions, this time rejecting any 2$\sigma$ outliers from the orbit. The orbit was then refitted with the reduced data set, and this process was repeated until no $2\sigma$ outliers remained. As a result, eight stars were rejected as outliers: three in latitude and five in velocity. This is consistent with the number of expected interlopers. We will return to the question of interlopers and their effects when we attempt to determine a possible progenitor in section 7. 

\subsection{Orbit Fitting Results}

\begin{figure}
\includegraphics[width=6.5in]{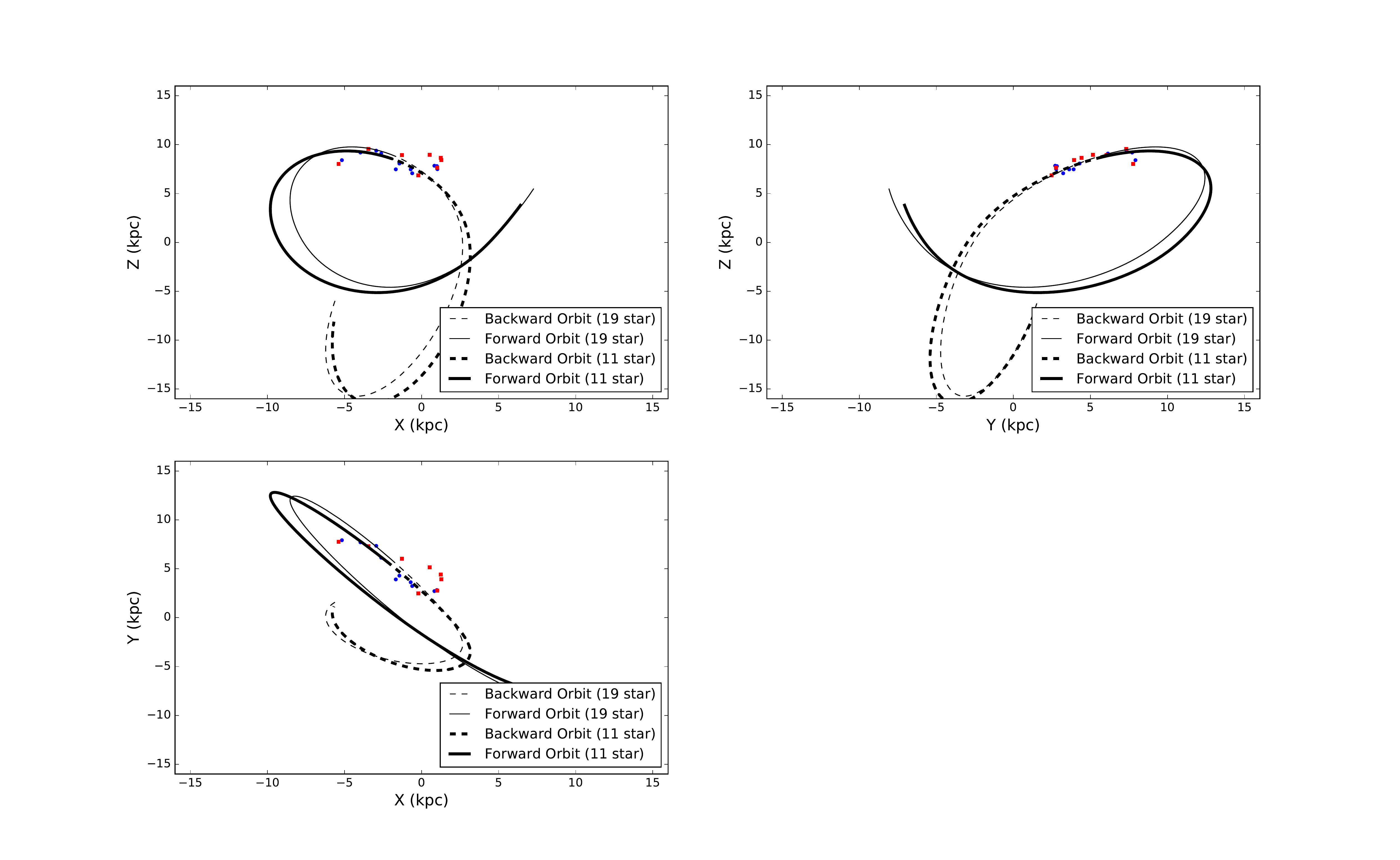}
\caption{Galactic XYZ plot of the eleven- and nineteen-star orbits and the data points. The eight removed as outliers are red. Each orbit is run for $250$ Myr forward and backward.}
\label{Hermus_XYZ_5}
\end{figure}

Table \ref{tab:Table2} gives the averaged best-fit parameters for a point on the orbit, using both the original nineteen stars and the eleven stars that remained after the outlier rejection process. Table \ref{tab:Table3} characterizes the orbits created from these initial points and evolved through the previously-stated potential. The two orbits along with the 19 stars are plotted in Figure \ref{Hermus_XYZ_5} in Galactic XYZ coordinates. The eight red squares represent the stars that were rejected using the outlier rejection process. These stars are also marked with an asterisk in the last column of Table \ref{tab:Table1}, which shows the deviation in v$_{\rm gsr}$ from the best-fit orbit. The left panels of Figure \ref{Hermus_lplots} show the stream parameters as a function of galactic longitude for the ten Hermus BHB candidates in comparison to the ten star best-fit orbit.

The metallicity of these 11 remaining stars is [Fe/H] $=-2.1 \pm 0.4$, including one star outside of the $2\sigma$ range. The metallicity becomes [Fe/H] $=-2.0 \pm 0.2$ if this outlier is removed. This metallicity is consistent with, but slightly higher than the measurement of [Fe/H]$\sim-2.3$ from G14.

\subsection{Stream Width}

To confirm that the detected moving group is narrow, like the Hermus Stream, and our selection of $\pm 2^\circ$ was reasonable, we made a histogram of the deviations from the best-fit orbit in RA. The top panels of Figure \ref{RA_diff} show histograms of SDSS BHB spectra with declinations $5^\circ<\delta<50^\circ$. The distance ranges were chosen to be consistent with the distances found in Figure \ref{vgsr_dist}, as well as the results of the best-fit orbit. There is a statistically significant ($3.0\sigma$) narrow peak that is centered at zero in both panels . Each bin is $2^\circ$ wide, which means that the initial selection of $\pm 2^\circ$ selected the full extent of the BHB structure. The bottom panels of Figure \ref{RA_diff} are the same as the top panels, but show the difference between the SDSS BHB spectra from the G14 third-order polynomial fit to Hermus. The excess around $0^\circ$ is more spread out. G14 created much narrower histograms of the deviations of photometrically selected stars from his fit to the Hermus Stream. However, he created these after removing a smooth background and after a median filter of five different regions along the stream. These operations could have introduced significant error into the width calculation, which might explain why his stream width of $0.^\circ 7$ is narrower than our measurement of $=1.^\circ6$ (calculated from the uncertainty of Galactic latitude of $\sigma_b = 0.68$). 

\subsection{Is this the Hermus Stream?}

The spatial coincidence of the Hermus Stream and our identified moving group suggests that they are the same structure. The alternative is that we detected a previously unidentified tidal stream that closely follows the path of the Hermus Stream across $50^\circ$ of the sky, and is within the distance range originally estimated in the G14 discovery paper on one end of the stream, but not the other. Since we searched the entire distance range in the discovery paper, the Hermus Stream, which has an estimated metallicity of [Fe/H]$=-2.3$, would have to be virtually devoid of BHB stars. Furthermore, G14 would have had to have missed this much closer stream in his search, in favor of the more distant stream. We have chosen to identify our stream with Hermus in part because distances derived from kernel-fitting have historically proven to be incorrect (see introduction).

\section{Age Estimates for Stream Stars}

We estimate ages for these stars following the procedure outlined in \citet{2015ApJ...813L..16S} and \citet{2016NatPh..12.1170C}. This method uses the colors and metallicities of BHB stars to determine their ages. Lower metallicity stars are bluer at a given age. The inclusion of metallicity in the age estimate improves the accuracy over previous methods. By selecting all spectra in the On and Off fields, and applying the same selection criteria outlined in the literature, ages for 369 stars were calculated. 

Figure \ref{Age_hist} shows the On (30 stars) and Off (50 stars) field stars between 10 and 14 kpc from the Sun; the Off field has been normalized to the number of On field stars. Of the 11 stars previously selected to belong to the Hermus Stream, eight of them had calculated ages, and are shown as the solid black histogram in Figure \ref{Age_hist}. Seven of these eight stars are found in the narrow age range of 11.3 to 12.8 Gyrs; the one outlier was the star previously identified as being an outlier in metallicity.

We tested whether the stream star ages were different from other BHBs in that region of the sky. Performing a two sample KS test of the seven stream stars with calculated ages and the 22 On stream (non-member) stars returns a p-value of $p=0.27$. If instead the entire 72 star sample is compared to the seven stream stars; a two sample KS test returns a p-value of $p=0.28$. 

The distribution of stream star ages seems consistent with being drawn from the same distribution as the rest of the BHB stars. The fact that seven of the eight stream stars occupy a narrow age range suggests that they could belong to a single structure. However, we have just confirmed that all of the BHB stars in the region seem to have a similar expected age and cannot be used to solely identify substructure in this region. Note that neither the ages nor the metallicities of the BHB stars were used to determine which stars were stream members.

\section{\textit{N}-body Simulation}

We simulated a possible progenitor of the Hermus Stream using the gyrfalcON \textit{N}-body integrator \citep{2000ApJ...536L..39D, 2002JCoPh.179...27D}. 
In order to constrain the progenitor mass, we performed \textit{N}-body simulations for average globular clusters ($10^5M_{\odot}, 10$ pc), large globular clusters \citep[$5\times10^5M_{\odot}, 15$ pc;][]{2008MNRAS.385L..20V}, and dwarf spheroidal (dSph) galaxies with [mass , scale radius] =  ($10^6M_{\odot}, 175$ pc), ($5\times10^6M_{\odot}, 250$ pc), ($10^7M_{\odot}, 500$ pc), ($5\times10^7M_{\odot}, 700$ pc), and ($10^8M_{\odot}, 800$ pc). The dSph galaxy masses and corresponding radii were selected from the mass-luminosity relation described in \citet{2009ApJ...704.1274W}. We also tried ultrafaint dwarf models with $10^5M_{\odot} < M < 5\times10^6M_{\odot}$ and scale radii of $20 \leq r \leq 50$ pc \citep{2016MNRAS.462.2734H}. 

The progenitor for each simulation was modeled with a single Plummer sphere profile, represented by $20,000$ particles. This is a valid representation for globular clusters, which are not expected to contain dark matter, so that mass follows light, and comparing the distribution of stream stars in the sky with the bodies in the simulation is justified. For dwarf galaxies, and especially for ultrafaint dwarfs that are believed to be heavily dominated by dark matter \citep{2015ApJ...813...44L, 2016MNRAS.458L..59M, 2015ApJ...814L...7K, 2016MNRAS.460.4397C}, the representation is less realistic. However, the simulations give some indication of the mass within the half-light radius. The simulated bodies are compared with the observed stars, but the width of the stream is influenced by the central concentration of dark matter in the dwarf galaxy progenitor.

As stated previously, the effects of interloper contamination, combined with the $\sim 10$ km s$^{-1}$ dispersion in duplicate measurements lead us to conclude that velocity dispersion is not a useful measure with which to characterize the progenitor. We instead focus on distance and angular dispersions to guide our choices for the progenitor's mass and scale radius.

We began by integrating each simulated dwarf galaxy radius/mass pair for $2$ Gyr along the orbit. The angular, velocity, and distance dispersions for each simulation result were then calculated for stars within $2^\circ$ of the orbit for $17^\circ < l < 72^\circ$ and $10 < $v$_{\rm gsr} < 110$ km s$^{-1}$, matching the criteria used to select the candidate BHBs

A comparison of the observed stream depth with the simulated stream depth allows us to rule out high-mass progenitors. The 11 BHB stars have a distance dispersion of $\sigma_r\sim0.80$ kpc from the orbit and an error in the distance calculation of $\sim1$ kpc. Since the dispersion in distance is similar to the measurement errors, the line-of-sight depth of the stream is consistent with zero, but could be as large as $1$ kpc if the calculated absolute magnitudes of the BHBs from \citet{2011MNRAS.416.2903D} have the same systematic offsets. The narrow distance dispersion rules out dSph progenitors of mass $10^7$, $5\times10^7$, or $10^8M_{\odot}$, which have simulated dispersions of $\sigma_r\sim1.24$, $\sigma_r\sim2.74$, and $\sigma_r\sim4.17$, respectively. The progenitors of mass $10^5$, $5\times10^5$, $10^6$ and $5\times10^6M_{\odot}$ are consistent with the data, with distance dispersions of $\sigma_r\sim0.06$, $\sigma_r\sim0.05$, $\sigma_r\sim0.42$ and $\sigma_r\sim0.87$, respectively. The distance dispersions favor a progenitor with a mass $<10^7M_{\odot}$.

We then investigated the dispersion in galactic latitude, $\sigma_{b}$. For our 11 BHBs, the angular dispersion is $\sigma_b=0.68^\circ$, giving us a full-width half max of $1.63^\circ$. We could reject the ($10^6M_{\odot}, 175$ pc), ($5\times10^6M_{\odot}, 250$ pc) as too wide, as their angular dispersions were $\sigma_b=1.09^\circ$ and $1.14^\circ$ respectively. To investigate further, we ran a new set of N-body simulations exploring various scale radii at masses of $5\times10^{5}M_{\odot}$ and $10^{6}M_{\odot}$. We found that one model, an ultrafaint dwarf with a mass of $10^6M_{\odot}$ and scale radius of $40$ pc, produced a dispersion ($\sigma_b=0.59^\circ$) possibly consistent with our observed angular dispersion. However, this model had a coherent core that does not appear in the data.

In order to further explore the parameter space of these ultrafaint progenitors, we then repeated runs of each scale radius/mass pair for 3 Gyr and 4 Gyr. We first evolved the orbit backwards for the appropriate amount of time to obtain a starting point, and then a Plummer sphere was evolved forward from that point for the corresponding time. For the progenitors of mass $M = 10^{5} M_{\odot}$ or $M = 5\times 10^{5} M_{\odot}$, we find that, despite the longer evolution times, they retain a coherent core along the stream which we do not see on the sky. To account for the coherent cores, we varied the forward evolution time of the Plummer sphere $\pm 50$ Myr for each backwards orbit evolution time ($2, 3,$ and $4$ Gyr). This moved the core off the sky location of our stream detection and allowed us to analyze how the leading and trailing tails would appear. For each model that had a core appear, we find that if the core is moved off the stream, the latitude dispersions of both the leading and trailing tails are too wide to fit our data. 

For the $M = 10^{6} M_{\odot}$ progenitors, we find that nearly all of them disrupt too much for our dispersion constraints. The one model that disrupts just enough to lack a core while remaining somewhat consistent with our dispersion results is the $10^6M_{\odot}$ and scale radius of $40$ pc evolved for $4$ Gyr. The dispersions of this simulation are $\sigma_b=0.86^\circ$, $\sigma_v=2.78^\circ$, and $\sigma_r=0.26^\circ$. The right panels of Figure \ref{Hermus_lplots} show \textit{N}-body results for the ultrafaint dwarf galaxy progenitor compared to the best-fit orbit. We evolved the progenitor for $4$ Gyr starting from $(X,Y,Z)=(-2.1,4.9,2.8)$ kpc with an initial velocity of $(v_x,v_y,v_z)=(-89,37,294)$ km s$^{-1}$ in Galactocentric coordinates with the Sun at $(-8,0,0)$ kpc. 

While we find that a dwarf of $(M,r_s) = (10^6M_{\odot},40$ pc) best fits our observations, we do not insist that it is the only possible progenitor. The parameter space is very large, and a comprehensive search of mass/scale radius combinations, including the case where mass does not follow light, is beyond the scope of this paper.

\section{No Association of Hermus with Other Known Substructure}

\begin{figure}
\includegraphics[width=6.5in]{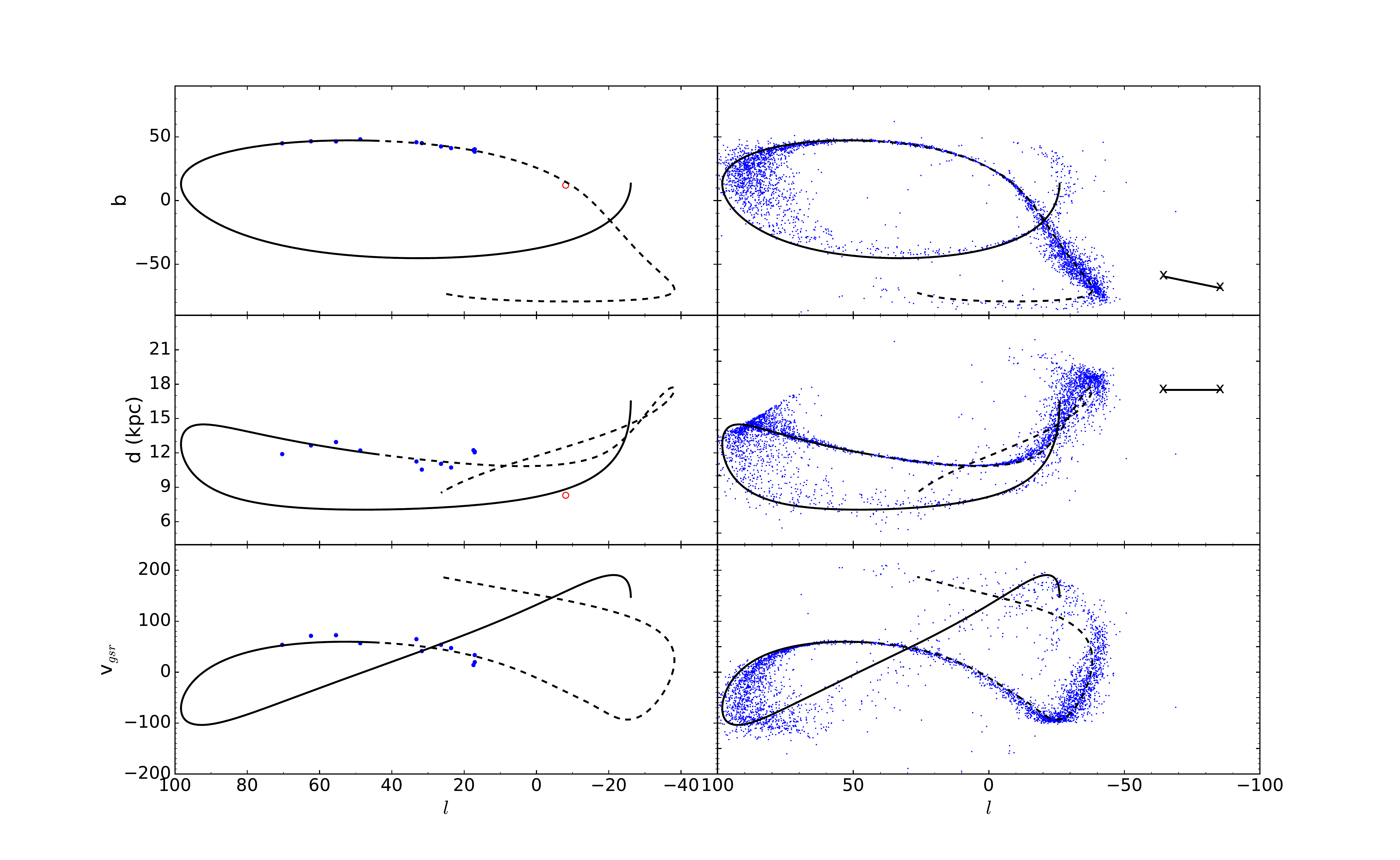}
\caption{Hermus eleven-star orbit sky position, distance, and line-of-sight velocity as a function of Galactic longitude. The forward (solid line) and backward (dashed line) orbits are shown for $250$ Myr from the orbit's starting point. Comparison to Figure 4 of G16 shows we find a similar position, but much closer distances. 
The left panels show the 11 Hermus BHB candidates plotted with the orbit while the right panels show a $20\%$ sub-sample of the \textit{N}-body results for a $10^6M_{\odot}$ progenitor of radius $40$ pc evolved $4$ Gyr along the best-fit orbit. The black x's mark the start and end positions of the previously observed position of the Phoenix Stream, which is apparently not associated with the Hermus Stream. Additionally, the parameters for globular cluster 1636-283 (red open circle) has been marked as a possible associated cluster.}
\label{Hermus_lplots}
\end{figure}

\begin{figure}
\includegraphics[width=6.5in]{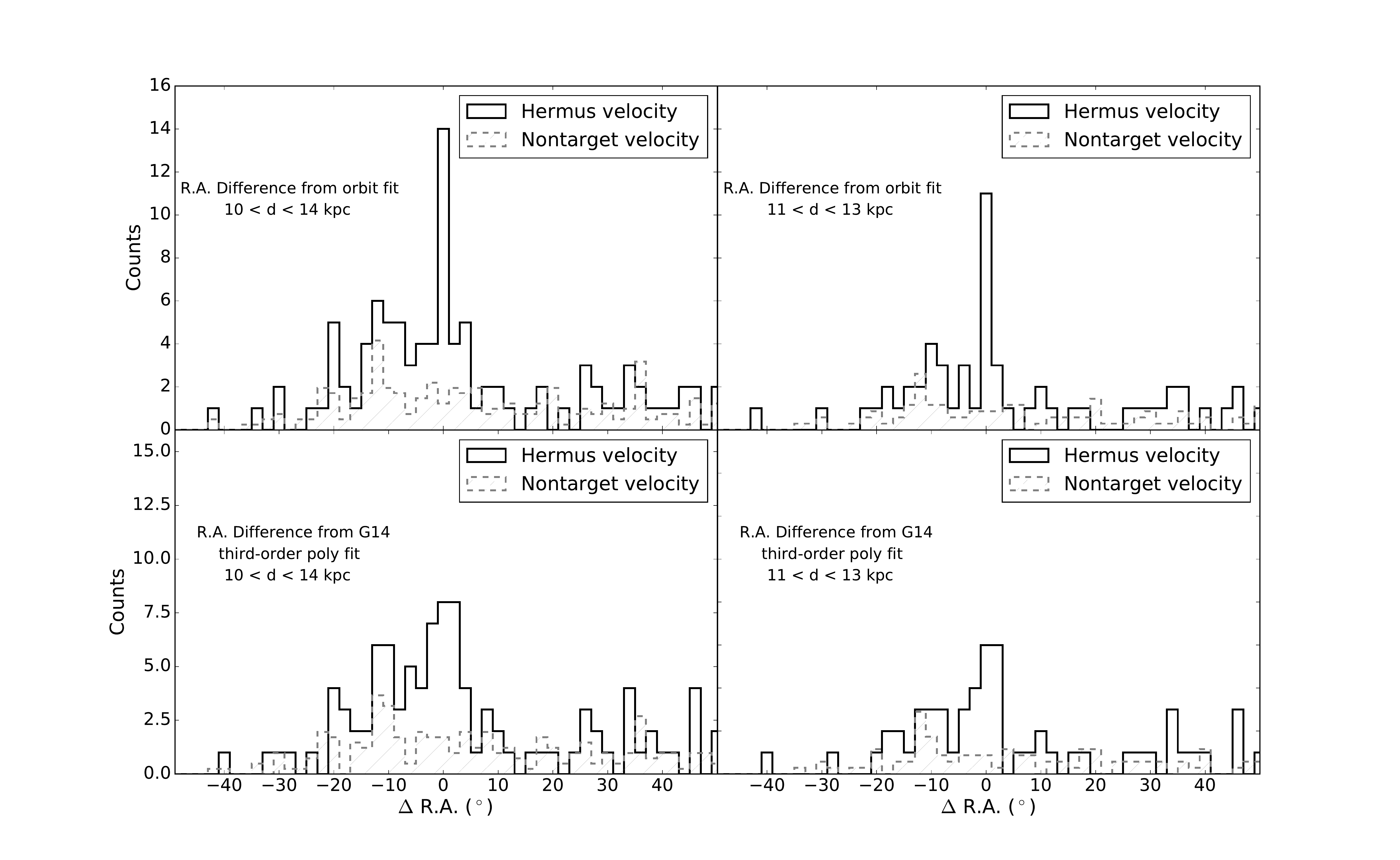}
\caption{Histogram of the difference in Right Accension (R.A.) of SDSS BHB spectra from the best-fit orbit (top panels) and the G14 third-order polynomial Hermus fit (bottom panels) in the range $5^\circ<\delta<50^\circ$. The open histograms represent all spectra within the Hermus velocity selection, while the hatched histograms represent all spectra outside the velocity selection, normalized to the number of Hermus velocity stars. The BHB distance range is given in each panel. {\it Top panels:} For both distance ranges there is a narrow ($2^\circ$) peak at $0^\circ$, both with $\sim 3.0\sigma$. {\it Bottom panels:} For both distances, the peak around $0^\circ$ is still evident but it is wider. This peak could plausibly be consistent with the narrow peak identified in G14.}
\label{RA_diff}
\end{figure}

\begin{figure}
\includegraphics[width=6.5in]{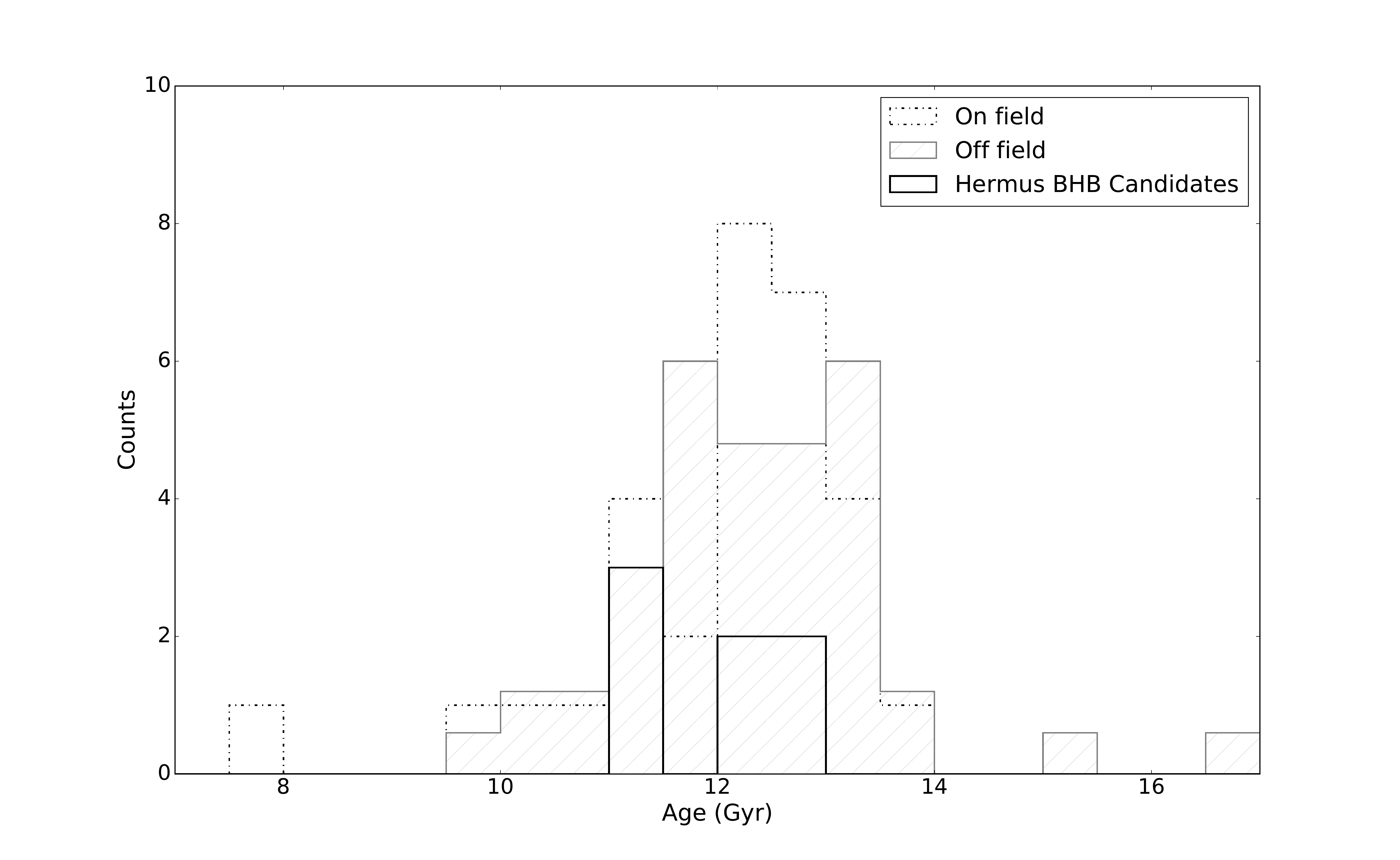}
\caption{Histogram of BHB stellar ages for stars with distances 10 to 14 kpc in the On (black dot-dashed histogram) and Off (gray hatched histogram) fields of Figure \ref{Hermus}. Seven of the eight Hermus stars with calculated ages are shown by the thick black histogram, the outlier star has been excluded which had an age of 7.76 Gyr. The Off field stars have been normalized to the number of On field stars.}
\label{Age_hist}
\end{figure}

Our orbit does not pass near the observed position of the Phoenix Stream in the sky, as shown in Figure \ref{Hermus_lplots}. The Phoenix Stream orbit was calculated by \citet{2016ApJ...820...58B} to be $18$ to $20$ kpc from the Galactic Center with a roughly circular orbit. This is in conflict with our finding that Hermus BHBs have a Galactocentric distance between $8$ and $12$ kpc along the entire orbit. Figure \ref{Hermus_lplots} shows that Galactocentric distances of $18$ to $20$ kpc are outside of the error box for the apogalacticon of our orbit fit. It is also inconsistent with the southern end of the Hermus Stream as identified by G14. Additionally, our orbit was compared to the positions of the newly discovered streams by the DES \citep{2018arXiv180103097S}, and we find no match. Our proposed orbit appears to travel perpendicular to several of the observed streams, or be at drastically different distances.

While the orbit we propose for the Hermus Stream does not link it with the Phoenix Stream, it does pass close ($\sim 4$ kpc) to the Galactic Center. This makes the Hermus Stream a potential candidate to probe the poorly known potential near the Galactic Center. 

In an attempt to identify possible satellite associations for the stream, we searched for globular clusters near (within $2^\circ$) the orbit using the Harris Catalog of globular clusters \citep[][2010 edition]{1996AJ....112.1487H}. We identified eight potential associated clusters based solely on their proximity to the path of the Hermus Stream's orbit. These eight objects are the globular clusters Palomar 14, 1636-283, NGC 6121, NGC, 6144, NGC 6352, NGC 6397, NGC 6752, and NGC 7099. We can immediately discard Palomar 14, due to its distance being well beyond that of the Hermus Stream. Likewise, the clusters NGC 6121, NGC 6352, NGC 6397, and NGC 6752 are all well within the Hermus Stream orbit by at least 9 kpc.  

The cluster 1636-283 is approximately $2$ kpc closer than the stream's orbit at its position and there is no published value for its radial velocity. Given the proximity of the globular cluster to the Galactic Center, we cannot rule out the possibility that this cluster could be associated with the stream. The position and distance for the cluster have been included as a red open circle in the left panels of Figure \ref{Hermus_lplots} for comparison to the orbit. 
For the remaining two objects, since they are at somewhat plausible heliocentric distances for association, we calculated a v$_{gsr}$ for each based on their listed radial velocities and positions, using the equation in Section 3. We then compared this radial velocity to the orbital radial velocity on our fitted orbit. We can reject both of these objects (NGC 6144 and NGC 7099), as their v$_{gsr}$ values are very different from the corresponding orbital v$_{gsr}$ values: NGC 6144 has a v$_{gsr}$ of $175$ km s$^{-1}$ compared to an orbital v$_{gsr}$ of $-35$ km s$^{-1}$, and NGC 7099 hs a v$_{gsr}$ of $-113$ km s$^{-1}$ compared to an orbital v$_{gsr}$ of $50$ km s$^{-1}$. 

We find that there is a single possible globular cluster that could be associated with our stream, 1636-283. Of particular note is that this cluster is classified as a "young halo" object \citep{2005MNRAS.360..631M, 2004MNRAS.355..504M}. Young halo objects are said to be consistent with being formed in satellite dwarf galaxy systems \citep{1993ASPC...48...38Z} that have been accreted by the Milky Way. Given that the stream appears to be consistent with an ultrafaint dwarf galaxy merger event, it is possible that this cluster could have originated in the dwarf galaxy progenitor and have since been stripped from the host. It is important to emphasize that we are only suggesting a possible association and not stating that it must be related. Observations of radial velocities for the cluster would be needed to confirm or rule out a possible association.

\section{Conclusion}
We find $19$ co-moving BHB stars at the sky position and approximate distance of the Hermus Stream. Of these, $11$ are stream members with high probability. An orbit fit to these stars can be described by a velocity of $\mathbf{v}_0= (-123, 155, 59)$ $\text{km}~\text{s}^{-1}$ at a position of $(X,Y,Z) = (-2.3, 5.7, 8.7)$ kpc. This orbit has a perigalacticon of $4.0$ kpc, apogalacticon of $17.0$ kpc, orbital period of $247$ Myr, orbital eccentricity $e = 0.62$, and orbital inclination $i=75.87^\circ$ with respect to the positive Z axis. 

The spatial coincidence of the Hermus Stream and our identified moving group suggests that they could be the same structure. The alternative is that we detected a previously unidentified tidal stream that closely follows the path of the Hermus Stream across $50^\circ$ of the sky, and is within the distance range originally estimated in the G14 discovery paper on one end of the stream, but not the other. Since we searched the entire distance range in the discovery paper, the Hermus Stream, which has an estimated metallicity of [Fe/H]$=-2.3$, would have to be virtually devoid of BHB stars. Furthermore, G14 would have had to have missed this much closer stream in his search, in favor of the more distant stream. 

The standard deviations from the orbit of the associated BHB stars' sky positions, line-of-sight velocities, and distances are $\sigma_b=0.68^\circ$, $\sigma_v=10.37$ km s$^{-1}$, and $\sigma_r=0.80$ kpc, respectively. Both the line-of-sight stream width and the width of the line-of-sight velocities could be considerably smaller, since the distance errors to the BHB stars are of order $1$ kpc ($\sim10\%$), and the radial velocity errors are likely to be $10$ km s$^{-1}$ or higher.

Seven of the eight potential Hermus Stream BHBs have ages withing the range 11.3-12.8 Gyr. The eigth star associated with the Hermus Stream in position and line-of-sight velocity is an outlier in metallicity, which is correlated with the age determination. The metallicity of the 11 BHB stars is [Fe/H] $=-2.1 \pm 0.4$, or $-2.0 \pm 0.2$ if the one outlier is removed.

The spatial distribution of these stars suggests a progenitor with mass $\lesssim10^7M_{\odot}$. We are able to match the spatial data for the Hermus Stream, and the spatial data for the 11 BHBs, with an ultrafaint dwarf galaxy of mass $10^{6}M_{\odot}$ and scale radius of $40$ pc, evolved for $4$ Gyr through a MW potential.

The stream is not associated with the Phoenix Stream. However, we identify a possibly associated young halo globular cluster, 1636-283, which may have formed in the ultrafaint dwarf galaxy progenitor before its merger with the Milky Way. The proposed orbit passes within $4$ kpc of the Galactic center, making it a candidate to probe the gravitational potential in this central region of the Milky Way.

\clearpage

\clearpage

\clearpage

\clearpage

\clearpage

\clearpage

\clearpage

\clearpage

\clearpage
\small
\begin{landscape}

\begin{table}
	\caption{Hermus Stream Candidate BHBs$\label{cpsk}$ }
	\label{tab:Table1}
	\begin{tabular}{lcccccccrccc} 
		\hline
		ra & dec & $l$ & $b$ &  $g_0$ & $(u-g)_0$ & $(g-r)_0$ & Dist$_{\rm Calc}$ & $v_{\rm gsr}$ & [Fe/H]$_{\rm WBG}$ & $\log g_{\rm WBG}$ & $v_{\rm gsr}$ Deviation\\
		($^\circ$, J2000) & ($^\circ$, J2000) & ($^\circ$) & ($^\circ$) &  &  &  & (kpc) & (km s$^{-1}$) &  &  & (km s$^{-1}$)\\
		\hline

240.4129 & 6.2931 & 17.075 & 40.317 & 16.05 & 1.17 & -0.24 & 12.12 & 19.57 & -1.94 & 3.21 & 11.538\\
240.4456 & 19.2716 & 33.213 & 45.819 & 15.70 & 1.18 & -0.05 & 11.25 & 64.77 & -3.11 & 1.94 & -12.976\\
240.7412 & 18.0466 & 31.709 & 45.136 & 15.66 & 1.17 & -0.20 & 10.54 & 41.38 & -1.77 & 3.41 & 9.093\\
240.8303 & 30.0871 & 48.743 & 48.087 & 16.08 & 1.09 & -0.24 & 12.21 & 56.83 & -2.47 & 3.05 & 2.798\\
240.8705 & 6.5133 & 17.625 & 40.036 & 15.60 & 1.17 & -0.10 & 10.66 & 107.32 & -1.41 & 3.12 & -75.235$^*$\\
241.1580 & 6.1691 & 17.425 & 39.616 & 15.95 & 1.20 & -0.17 & 12.24 & 14.03 & -1.95 & 2.81 & 17.704\\
241.5433 & 5.6415 & 17.090 & 39.018 & 15.96 & 1.16 & -0.20 & 12.14 & 88.85 & -1.88 & 3.30 & -57.714$^*$\\
241.6348 & 36.3307 & 58.136 & 47.958 & 16.02 & 1.24 & -0.13 & 12.87 & 93.44 & -1.98 & 2.90 & -33.932$^*$\\
241.7193 & 13.5146 & 26.411 & 42.522 & 15.74 & 1.21 & -0.19 & 11.05 & 53.75 & -2.00 & 2.87 & -8.830\\
242.0014 & 11.0777 & 23.617 & 41.223 & 15.62 & 1.18 & -0.11 & 10.73 & 47.29 & -1.71 & 2.98 & -5.924\\
242.0710 & 5.3371 & 17.097 & 38.411 & 15.95 & 1.17 & -0.20 & 12.05 & 33.52 & -1.92 & 3.02 & -2.371\\
243.0244 & 9.9939 & 22.942 & 39.836 & 16.10 & 1.18 & -0.17 & 13.13 & 101.46 & -2.16 & 2.79 & -61.025$^*$\\
243.3505 & 34.4308 & 55.458 & 46.455 & 16.02 & 1.18 & -0.10 & 12.95 & 72.51 & -2.09 & 2.32 & -12.575\\
243.4221 & 39.1946 & 62.381 & 46.578 & 16.03 & 1.19 & -0.18 & 12.65 & 71.27 & -2.16 & 3.46 & -13.119\\
243.5326 & 24.8021 & 41.916 & 44.691 & 15.96 & 1.10 & -0.01 & 12.70 & 49.95 & -1.14 & 3.46 & 7.477$^*$\\
243.6432 & 11.9341 & 25.563 & 40.153 & 16.07 & 1.08 & 0.00 & 13.41 & 20.36 & -2.38 & 3.42 & 23.530$^*$\\
243.6772 & 16.5176 & 31.147 & 41.969 & 16.19 & 1.13 & -0.20 & 13.40 & 91.09 & -1.69 & 3.22 & -41.134$^*$\\
245.2142 & 44.8072 & 70.338 & 44.992 & 15.82 & 1.21 & -0.05 & 11.90 & 53.58 & -1.65 & 2.25 & -0.685\\
245.9134 & 45.6098 & 71.403 & 44.432 & 15.83 & 1.27 & -0.19 & 11.48 & 33.25 & -1.74 & 2.87 & 18.620$^*$\\
		\hline
	\end{tabular}
\\
$^*$ Removed as a $>2\sigma$ outlier in velocity or latitude.
\end{table}

\begin{table}
	
	\caption{Average Orbit Fit Results for 11 and 19 Stars}
	\label{tab:Table2}
	\begin{tabular}{lll} 
		\hline
		 & 11 star orbit & 15 star orbit \\
		\hline 

$l (^\circ)$ & $45$ & $45$ \\
$\overline{b}$ $(^\circ)$ & $47.05 \pm 0.06$ & $46.84 \pm 0.06$ \\
$\overline{d}$ (kpc)  & $11.92\pm 0.15 $  & $12.38 \pm 0.15$ \\
$\overline{v}_{x}$ (km s$^{-1})$ & $-122.82 \pm 4.01$ & $-107.00 \pm 2.82$ \\
$\overline{v}_{y}$ (km s$^{-1})$ & $154.81 \pm 4.45$ & $146.78 \pm 3.23$ \\
$\overline{v}_{z}$ (km s$^{-1})$ & $58.84 \pm 1.98$ & $64.30 \pm 1.67$ \\
		\hline
	\end{tabular}
\end{table}

\begin{table}
	
	\caption{Average Fitted Orbit Parameters}
	\label{tab:Table3}
	\begin{tabular}{lll}
		\hline
		& 11 star orbit & 19 star orbit \\
		\hline

$r_{p}$ (kpc) & 3.96 & 3.52 \\
$r_{a}$ (kpc)& 16.98 & 16.61  \\
$e$ & 0.62 & 0.65 \\
$i~ (^\circ$ w.r.t pos. $z$-axis) & 75.87 & 75.31 \\
$T ~$(Myr) &247 & 236 \\
		\hline
	\end{tabular}
\end{table}

\end{landscape}

\section*{Acknowledgements}

We thank the anonymous referee for insightful comments. We acknowledge support from NSF AST 16-15688, the NASA/NY Space Grant fellowship, and contributions made by The Marvin Clan, Babette Josephs, Manit Limlamai, and the 2015 Crowd Funding Campaign to Support Milky Way Research. JLC acknowledges support from AST 11-51462. TCB acknowledges partial support for this work from grant PHY 14-30152; Physics Frontier Center/JINA Center for the Evolution of the Elements (JINA-CEE), awarded by the US National Science Foundation. PD acknowledges partial funding from a Natural Sciences and Engineering Research Council of Canada grant to D. VandenBerg. Funding for SDSS-III has been provided by the Alfred P. Sloan Foundation, the Participating Institutions, the National Science Foundation, and the US Department of Energy Office of Science. The SDSS-III Website is http://www.sdss3.org/.

\bsp	
\label{lastpage}

\begin{thebibliography}{99}

\bibitem[Ahn et al.(2014)]{2014ApJS..211...17A} Ahn, C.~P., Alexandroff, R., Allende Prieto, C., et al.\ 2014, \apjs, 211, 17

\bibitem[Anderson \& Darling (1952)]{AndersonDarling} Anderson, T. W.; Darling, D. A. Asymptotic Theory of Certain "Goodness of Fit" Criteria Based on Stochastic Processes. Ann. Math. Statist. 23 (1952), no. 2, 193--212.

\bibitem[Balbinot et al.(2016)]{2016ApJ...820...58B} Balbinot, E., Yanny, B., Li, T.~S., et al.\ 2016, \apj, 820, 58 

\bibitem[Belokurov et al.(2006)]{2006ApJ...642L.137B} Belokurov, V., 
Zucker, D.~B., Evans, N.~W., et al.\ 2006, \apjl, 642, L137

\bibitem[Belokurov et al.(2007a)]{2007ApJ...657L..89B} Belokurov, V., Evans, 
N.~W., Bell, E.~F., et al.\ 2007a, \apjl, 657, L89 

\bibitem[Belokurov et al.(2007b)]{2007ApJ...658..337B} Belokurov, V., Evans, 
N.~W., Irwin, M.~J., et al.\ 2007b, \apj, 658, 337 

\bibitem[Belokurov(2013)]{2013NewAR..57..100B} Belokurov, V.\ 2013, NewAR, 
57, 100 

\bibitem[Bernard et al.(2014)]{2014MNRAS.443L..84B} Bernard, E.~J., Ferguson, A.~M.~N., Schlafly, E.~F., et al.\ 2014, \mnras, 443, L84 

\bibitem[Bernard et al.(2016)]{2016MNRAS.463.1759B} Bernard, E.~J., Ferguson, A.~M.~N., Schlafly, E.~F., et al.\ 2016, \mnras, 463, 1759 

\bibitem[Bonaca et al.(2012)]{2012ApJ...760L...6B} Bonaca, A., Geha, M., \& Kallivayalil, N.\ 2012, \apjl, 760, L6 

\bibitem[Bonaca et al.(2014)]{2014ApJ...795...94B} Bonaca, A., Geha, M., K{\"u}pper, A.~H.~W., et al.\ 2014, \apj, 795, 94 

\bibitem[Bovy(2016)]{2016PhRvL.116l1301B} Bovy, J.\ 2016, Physical Review Letters, 116, 121301 

\bibitem[Bovy et al.(2016)]{2016ApJ...833...31B} Bovy, J., Bahmanyar, A., Fritz, T.~K., \& Kallivayalil, N.\ 2016, \apj, 833, 31 

\bibitem[Brown et al.(2005)]{2005ApJ...622L..33B} Brown, W.~R., Geller, 
M.~J., Kenyon, S.~J., \& Kurtz, M.~J.\ 2005, \apjl, 622, L33

\bibitem[Calabrese \& Spergel(2016)]{2016MNRAS.460.4397C} Calabrese, E., \& Spergel, D.~N.\ 2016, \mnras, 460, 4397 

\bibitem[Carlberg(2009)]{2009ApJ...705L.223C} Carlberg, R.~G.\ 2009, \apjl, 705, L223 

\bibitem[Carlberg \& Grillmair(2013)]{2013ApJ...768..171C} Carlberg, R.~G., \& Grillmair, C.~J.\ 2013, \apj, 768, 171 

\bibitem[Carlin et al.(2012a)]{2012ApJ...744...25C} Carlin, J.~L., Majewski, S.~R., Casetti-Dinescu, D.~I., et al.\ 2012a, \apj, 744, 25 

\bibitem[Carlin et al.(2012b)]{2012ApJ...753..145C} Carlin, J.~L., Yam, W., 
Casetti-Dinescu, D.~I., et al.\ 2012b, \apj, 753, 145

\bibitem[Carollo et al.(2016)]{2016NatPh..12.1170C} Carollo, D., Beers, T.~C., Placco, V.~M., et al.\ 2016, Nature Physics, 12, 1170 

\bibitem[Crnojevi{\'c}(2016)]{2016arXiv161205471C} Crnojevi{\'c}, D.\ 2016, arXiv:1612.05471 

\bibitem[D'Agostino \& Stephens(1986)]{1986gft..book.....D} D'Agostino, R.~B., \& Stephens, M.~A.\ 1986, Statistics: Textbooks and Monographs, New York: Dekker, 1986, edited by D'Agostino, Ralph B.; Stephens, Michael A.,  


\bibitem[Deason et al.(2011)]{2011MNRAS.416.2903D} Deason, A.~J., 
Belokurov, V., \& Evans, N.~W.\ 2011, \mnras, 416, 2903

\bibitem[Dehnen(2000)]{2000ApJ...536L..39D} Dehnen, W.\ 2000, \apjl, 536, 
L39 

\bibitem[Dehnen(2002)]{2002JCoPh.179...27D} Dehnen, W.\ 2002, Journal of 
Computational Physics, 179, 27 

\bibitem[Dierickx \& Loeb(2017)]{2017arXiv170302137D} Dierickx, M.~I.~P., \& Loeb, A.\ 2017, arXiv:1703.02137 

\bibitem[Duffau et al.(2006)]{2006ApJ...636L..97D} Duffau, S., Zinn, R., 
Vivas, A.~K., et al.\ 2006, \apjl, 636, L97 

\bibitem[Duffau et 
al.(2014)]{2014A&A...566A.118D} Duffau, S., Vivas, A.~K., Zinn, R., M{\'e}ndez, R.~A., \& Ruiz, M.~T.\ 2014, \aap, 566, AA118 

\bibitem[Dumas et al.(2015)]{2015ApJ...811...36D} Dumas, J., Newberg, H.~J., Niedzielski, B., et al.\ 2015, \apj, 811, 36

\bibitem[Erkal et al.(2016)]{2016MNRAS.463..102E} Erkal, D., Belokurov, V., Bovy, J., \& Sanders, J.~L.\ 2016, \mnras, 463, 102 

\bibitem[Gibbons et al.(2014)]{2014MNRAS.445.3788G} Gibbons, S.~L.~J., 
Belokurov, V., \& Evans, N.~W.\ 2014, \mnras, 445, 3788 

\bibitem[Gibbons et al.(2016)]{2016MNRAS.458L..64G} Gibbons, S.~L.~J., Belokurov, V., Erkal, D., \& Evans, N.~W.\ 2016, \mnras, 458, L64 

\bibitem[Grillmair 
\& Dionatos(2006)]{2006ApJ...643L..17G} Grillmair, C.~J., \& Dionatos, O.\ 2006, \apjl, 643, L17 

\bibitem[Grillmair \& Johnson(2006)]{2006ApJ...639L..17G} Grillmair, C.~J., \& Johnson, R.\ 2006, \apjl, 639, L17

\bibitem[Grillmair(2006a)]{2006ApJ...645L..37G} Grillmair, C.~J.\ 2006a, \apjl, 645, L37 

\bibitem[Grillmair(2006b)]{2006ApJ...651L..29G} Grillmair, C.~J.\ 2006b, \apjl, 651, L29

\bibitem[Grillmair(2009)]{2009ApJ...693.1118G} Grillmair, C.~J.\ 2009, 
\apj, 693, 1118

\bibitem[Grillmair(2012)]{2012ASPC..458..219G} Grillmair, C.~J.\ 2012, Galactic Archaeology: Near-Field Cosmology and the Formation of the Milky Way, 458, 219 

\bibitem[Grillmair et al.(2013)]{2013ApJ...769L..23G} Grillmair, C.~J., Cutri, R., Masci, F.~J., et al.\ 2013, \apjl, 769, L23 

\bibitem[Grillmair(2014)]{2014ApJ...790L..10G} Grillmair, C.~J.\ 2014, \apjl, 790, LL10 

\bibitem[Grillmair \& Carlberg(2016)]{2016ApJ...820L..27G} Grillmair, C.~J., \& Carlberg, R.~G.\ 2016, \apjl, 820, L27 

\bibitem[Grillmair 
\& Carlin(2016)]{2016ASSL..420...87G} Grillmair, C.~J., \& Carlin, J.~L.\ 2016, Astrophysics and Space Science Library, 420, 87 

\bibitem[Grillmair(2017)]{2017ApJ...834...98G} Grillmair, C.~J.\ 2017, \apj, 834, 98 

\bibitem[Gross \& Vitells(2010)]{2010EPJC...70..525G} Gross, E., \& Vitells, O.\ 2010, European Physical Journal C, 70, 525.

\bibitem[Harris(1996)]{1996AJ....112.1487H} Harris, W.~E.\ 1996, \aj, 112, 1487 

\bibitem[Helmi \& White(1999)]{1999MNRAS.307..495H} Helmi, A., \& White, S.~D.~M.\ 1999, \mnras, 307, 495

\bibitem[Helmi et al.(2017)]{2017A&A...598A..58H} Helmi, A., Veljanoski, J., Breddels, M.~A., Tian, H., \& Sales, L.~V.\ 2017, \aap, 598, A58  

\bibitem[Hernandez(2016)]{2016MNRAS.462.2734H} Hernandez, X.\ 2016, \mnras, 462, 2734 

\bibitem[Ibata et al.(2001a)]{2001ApJ...547L.133I} Ibata, R., Irwin, M., 
Lewis, G.~F., \& Stolte, A.\ 2001a, \apjl, 547, L133

\bibitem[Ibata et al.(2001b)]{2001ApJ...551..294I} Ibata, R., Lewis, G.~F., Irwin, M., Totten, E., \& Quinn, T.\ 2001b, \apj, 551, 294 

\bibitem[Johnston et al.(2005)]{2005ApJ...619..800J} Johnston, K.~V., Law, D.~R., \& Majewski, S.~R.\ 2005, \apj, 619, 800 

\bibitem[Kepley et al.(2007)]{2007AJ....134.1579K} Kepley, A.~A., Morrison, H.~L., Helmi, A., et al.\ 2007, \aj, 134, 1579 

\bibitem[Kirby et al.(2015)]{2015ApJ...814L...7K} Kirby, E.~N., Cohen, J.~G., Simon, J.~D., \& Guhathakurta, P.\ 2015, \apjl, 814, L7 

\bibitem[Kollmeier et al.(2009)]{2009ApJ...705L.158K} Kollmeier, J.~A., 
Gould, A., Shectman, S., et al.\ 2009, \apjl, 705, L158 

\bibitem[Koposov et al.(2010)]{2010ApJ...712..260K} Koposov, S.~E., Rix, 
H.-W., \& Hogg, D.~W.\ 2010, \apj, 712, 260 

\bibitem[Koposov et al.(2014)]{2014MNRAS.442L..85K} Koposov, S.~E., Irwin, M., Belokurov, V., et al.\ 2014, \mnras, 442, L85 

\bibitem[Laevens et al.(2015)]{2015ApJ...813...44L} Laevens, B.~P.~M., Martin, N.~F., Bernard, E.~J., et al.\ 2015, \apj, 813, 44 

\bibitem[Law et al.(2005)]{2005ApJ...619..807L} Law, D.~R., Johnston, K.~V., \& Majewski, S.~R.\ 2005, \apj, 619, 807 

\bibitem[Law \& Majewski(2010)]{2010ApJ...714..229L} Law, D.~R., \& Majewski, S.~R.\ 2010, \apj, 714, 229 

\bibitem[Li et al.(2016)]{2016ApJ...817..135L} Li, T.~S., Balbinot, E., Mondrik, N., et al.\ 2016, \apj, 817, 135 

\bibitem[Mackey \& Gilmore(2004)]{2004MNRAS.355..504M} Mackey, A.~D., \& Gilmore, G.~F.\ 2004, \mnras, 355, 504 

\bibitem[Mackey \& van den Bergh(2005)]{2005MNRAS.360..631M} Mackey, A.~D., \& van den Bergh, S.\ 2005, \mnras, 360, 631 

\bibitem[Mahalanobis (1936)]{Mahalanobis} Mahalanobis, P.~C., On the generalised distance in statistics. 
{\it Proceedings of the National Institute of Sciences of India} 1936 {\bf 2}. 49-55 

\bibitem[Majewski et al.(2003)]{2003ApJ...599.1082M} Majewski, S.~R., 
Skrutskie, M.~F., Weinberg, M.~D., 
\& Ostheimer, J.~C.\ 2003, \apj, 599, 1082 

\bibitem[Martin et al.(2013)]{2013ApJ...765L..39M} Martin, C., Carlin, 
J.~L., Newberg, H.~J., \& Grillmair, C.\ 2013, \apjl, 765, LL39 

\bibitem[Martin et al.(2014)]{2014ApJ...787...19M} Martin, N.~F., Ibata, R.~A., Rich, R.~M., et al.\ 2014, \apj, 787, 19 

\bibitem[Martin et al.(2016)]{2016MNRAS.458L..59M} Martin, N.~F., Geha, M., Ibata, R.~A., et al.\ 2016, \mnras, 458, L59 

\bibitem[Newberg et al.(2002)]{2002ApJ...569..245N} Newberg, H.~J., Yanny, 
B., Rockosi, C., et al.\ 2002, \apj, 569, 245 

\bibitem[Newberg et al.(2007)]{2007ApJ...668..221N} Newberg, H.~J., Yanny, 
B., Cole, N., et al.\ 2007, \apj, 668, 221 

\bibitem[Newberg et al.(2009)]{2009ApJ...700L..61N} Newberg, H.~J., Yanny, 
B., \& Willett, B.~A.\ 2009, \apjl, 700, L61 

\bibitem[Newberg et al.(2010)]{2010ApJ...711...32N} Newberg, H.~J., 
Willett, B.~A., Yanny, B., \& Xu, Y.\ 2010, \apj, 711, 32

\bibitem[Newberg \& Carlin(2016)]{2016ASSL..420.....N} Newberg, H.~J., \& Carlin, J.~L.\ 2016, Astrophysics and Space Science Library, 420,  

\bibitem[Odenkirchen et al.(2001)]{2001ApJ...548L.165O} Odenkirchen, M., Grebel, E.~K., Rockosi, C.~M., et al.\ 2001, \apjl, 548, L165 

\bibitem[Price-Whelan \& Johnston(2013)]{2013ApJ...778L..12P} Price-Whelan, A.~M., \& Johnston, K.~V.\ 2013, \apjl, 778, L12 

\bibitem[Rockosi et al.(2002)]{2002AJ....124..349R} Rockosi, C.~M., Odenkirchen, M., Grebel, E.~K., et al.\ 2002, \aj, 124, 349 

\bibitem[Sanders \& Binney(2013)]{2013MNRAS.433.1813S} Sanders, J.~L., \& Binney, J.\ 2013, \mnras, 433, 1813

\bibitem[Sanderson et al.(2015)]{2015ApJ...801...98S} Sanderson, R.~E., Helmi, A., \& Hogg, D.~W.\ 2015, \apj, 801, 98 

\bibitem[Sanderson et al.(2017)]{2017ApJ...836..234S} Sanderson, R.~E., Hartke, J., \& Helmi, A.\ 2017, \apj, 836, 234 

\bibitem[Santucci et al.(2015)]{2015ApJ...813L..16S} Santucci, R.~M., Beers, T.~C., Placco, V.~M., et al.\ 2015, \apjl, 813, L16

\bibitem[Sesar et al.(2007)]{2007AJ....134.2236S} Sesar, B., Ivezi{\'c}, 
{\v Z}., Lupton, R.~H., et al.\ 2007, \aj, 134, 2236

\bibitem[Sesar et al.(2010)]{2010ApJ...717..133S} Sesar, B., Vivas, A.~K., 
Duffau, S., \& Ivezi{\'c}, {\v Z}.\ 2010, \apj, 717, 133 

\bibitem[Sesar et al.(2013)]{2013ApJ...776...26S} Sesar, B., Grillmair, 
C.~J., Cohen, J.~G., et al.\ 2013, \apj, 776, 26 

\bibitem[Shipp et al.(2018)]{2018arXiv180103097S} Shipp, N., Drlica-Wagner, A., Balbinot, E., et al.\ 2018, arXiv:1801.03097 

\bibitem[Simion et al.(2014)]{2014MNRAS.440..161S} Simion, I.~T., 
Belokurov, V., Irwin, M., \& Koposov, S.~E.\ 2014, \mnras, 440, 161

\bibitem[Starkenburg et al.(2009)]{2009ApJ...698..567S} Starkenburg, E., Helmi, A., Morrison, H.~L., et al.\ 2009, \apj, 698, 567 

\bibitem[Teuben(1995)]{1995ASPC...77..398T} Teuben, P.\ 1995, Astronomical 
Data Analysis Software and Systems IV, 77, 398 

\bibitem[van den Bergh(2008)]{2008MNRAS.385L..20V} van den Bergh, S.\ 2008, \mnras, 385, L20

\bibitem[Vera-Ciro \& Helmi(2013)]{2013ApJ...773L...4V} Vera-Ciro, C., \& Helmi, A.\ 2013, \apjl, 773, L4 

\bibitem[Vivas et al.(2001)]{2001ApJ...554L..33V} Vivas, A.~K., Zinn, R., 
Andrews, P., et al.\ 2001, \apjl, 554, L33

\bibitem[Vivas et al.(2011)]{2011RMxAC..40..261V} Vivas, A.~K., Sesar, B., 
Duffau, S., 
\& Ivezic, Z.\ 2011, Revista Mexicana de Astronomia y Astrofisica Conference Series, 40, 261 

\bibitem[Walker et al.(2009)]{2009ApJ...704.1274W} Walker, M.~G., Mateo, M., Olszewski, E.~W., et al.\ 2009, \apj, 704, 1274 

\bibitem[Wilhelm et al.(1999)]{1999AJ....117.2329W} Wilhelm, R., Beers, 
T.~C., Sommer-Larsen, J., et al.\ 1999, \aj, 117, 2329

\bibitem[Willett et al.(2009)]{2009ApJ...697..207W} Willett, B.~A., 
Newberg, H.~J., Zhang, H., Yanny, B., \& Beers, T.~C.\ 2009, \apj, 697, 207

\bibitem[Willett(2010)]{2010PhDT.......194W} Willett, B.~A.\ 2010, Ph.D.~Thesis,  

\bibitem[Yam et al.(2013)]{2013ApJ...776..133Y} Yam, W., Carlin, J.~L., 
Newberg, H.~J., et al.\ 2013, \apj, 776, 133 

\bibitem[Yanny et al.(2000)]{2000ApJ...540..825Y} Yanny, B., Newberg, 
H.~J., Kent, S., et al.\ 2000, \apj, 540, 825

\bibitem[Yanny et al.(2009)]{2009AJ....137.4377Y} Yanny, B., Rockosi, C., Newberg, H.~J., et al.\ 2009, \aj, 137, 4377-4399 

\bibitem[York et al.(2000)]{2000AJ....120.1579Y} York, D.~G., Adelman, J., Anderson, J.~E., Jr., et al.\ 2000, \aj, 120, 1579 

\bibitem[Zinn(1993)]{1993ASPC...48...38Z} Zinn, R.\ 1993, The Globular Cluster-Galaxy Connection, 48, 38 

\end{thebibliography}
\end{document}